\documentclass[amsmath,amssymb]{revtex4-2}
\usepackage{slashed}
\usepackage{graphics}
\usepackage{epsfig}
\usepackage{array}
\usepackage{float}
\usepackage{xcolor}
\usepackage{colortbl}
\usepackage{booktabs}
\usepackage{graphicx}
\usepackage{psfrag}
\usepackage{slashbox}
\usepackage{hyperref}
\usepackage[normalem]{ulem}
\def\lsim{\raise0.3ex\hbox{$\;<$\kern-0.75em\raise-1.1ex\hbox{$\sim\;$}}}
\def\gsim{\raise0.3ex\hbox{$\;>$\kern-0.75em\raise-1.1ex\hbox{$\sim\;$}}}

\def\ie{{\it i.e.,~}}
\newcommand{\be}{\begin{equation}}
\newcommand{\ee}{\end{equation}}

\newcommand{\ufo}{{\sc UFO}}

\newcommand{\madgraph}{{\sc MadGraph}}

\newcommand{\madanalysis}{{\sc MadAnalysis}}

\newcommand{\sarah}{{\sc Sarah}}
\newcommand{\spheno}{{\sc SPheno}}
\newcommand{\higgsbounds}{{\sc HiggsBounds}}
\newcommand{\higgssignals}{{\sc HiggsSignals}}

\newcommand{\pythia}{{\sc Pythia}}
\newcommand{\delphes}{{\sc Delphes}}

\newcommand{\montecarlo}{{\sc MonteCarlo}}

\makeatletter

 \def\CT@@do@color{%
 \global\let\CT@do@color\relax
 \@tempdima\wd\z@
 \advance\@tempdima\@tempdimb
 \advance\@tempdima\@tempdimc
 \advance\@tempdimb\tabcolsep
 \advance\@tempdimc\tabcolsep
 \advance\@tempdima2\tabcolsep
 \kern-\@tempdimb
 \leaders\vrule
 \hskip\@tempdima\@plus 1fill
 \kern-\@tempdimc
 \hskip-\wd\z@ \@plus -1fill }
 \makeatother
\begin{document}
\title{Search for a heavy neutral Higgs boson in a left-right model with an inverse seesaw mechanism at the LHC}
\author{
M. Ashry$^{1,3,}$\footnote{mustafa@sci.cu.edu.eg}, K. Ezzat$^{2,3,}$\footnote{kareemezat@sci.asu.edu.eg} and S. Khalil$^{3,}$\footnote{skhalil@zewailcity.edu.eg}}
\affiliation{
$^1$Department of Mathematics, Faculty of Science, Cairo University, Giza 12613, Egypt.\\
$^2$Department of Mathematics, Faculty of Science, Ain Shams University, Cairo 11566, Egypt.\\
$^3$Center for Fundamental Physics, Zewail City of Science and Technology, 6th of October City, Giza 12578, Egypt.}
\date{\today}

\begin{abstract}
We develop a low scale left-right symmetric model based on $SU(3)_C\times SU(2)_L\times SU(2)_R\times U(1)_{B-L}\times Z_2$ with a simplified Higgs sector consisting of only one bidoublet and one $SU(2)_R$ doublet. In this model, the tiny values of light neutrino masses are generated through an inverse-seesaw mechanism. We emphasize that in this setup, the tree-level flavor changing neutral current can be strongly suppressed, consistent with the current experimental constraints. We show that the lightest $CP$-even Higgs boson, which is like the standard model Higgs boson, and the next lightest Higgs boson, $h'$, are generated from the neutral components of the bidoublet. We show that the mass of the next lightest Higgs boson can be of an order a few hundred~\text{GeVs}. 
We analyze the detection of $h'$ at the Large Hadron Collider (LHC) for a center-of-mass energy $\sqrt{s}=14~\text{TeV}$ and integrated luminosity $L_{\text{int}}=300~{\text{fb}}^{-1}$ via di-Higgs channel: $h'\to hh\to b\bar{b}\gamma\gamma$ and also in the $ZZ$ channel: $h'\to ZZ\to 4\ell~(\ell=e,~\mu)$ at an integrated luminosity $L_{\text{int}}=3000~{\text{fb}}^{-1}$. We consider three benchmark points for this analysis with $m_{h'}=250~\text{GeV},~400~\text{GeV}$, and $600~\text{GeV}$. 
We show that promising signals with good statistical significances can be obtained in di-Higgs channel, with $2\gamma+2b$-jets final states.
\end{abstract}
\maketitle

\section{Introduction}
The Standard Model (SM) of particle physics is in an excellent agreement with most of the confirmed experimental results. However, there exist several compelling arguments that indicate that the SM is only an effective low energy limit of a more fundamental underlying theory. Indeed, there are a number of theoretical and phenomenological outstanding issues in particle physics that can not be explained and the SM fails to address them adequately. Here, we may just mention the puzzles of dark matter and tiny neutrino masses~\cite{Fukuda:1998mi,Ahn:2002up,An:2012eh,Eguchi:2002dm,Ahmad:2002jz}, which can not be explained within the SM. One of the most popular extensions of the SM is the grand unified theory (GUT), where the SM gauge symmetry $SU(3)_C\times SU(2)_L\times U(1)_{Y}$ is extended to a bigger (simple or semisimple) group. 
Nonvanishing neutrino masses motivate the existence of right-handed neutrinos, and hence, all known fermions would have both left and right chirality. In this respect, the SM gauge group would be extended to the left-right (LR) symmetric gauge group, which is based on $SU(3)_C\times SU(2)_L\times SU(2)_R\times U(1)_{B-L}$, where left and right chirality are treated equally at high energy scales. In this class of models, the Majorana right-handed neutrinos are naturally heavy, and hence small left-handed neutrino masses are generated through seesaw mechanisms. 

In the conventional LR model proposed by Mohapatra {\it et al.}~\cite{Mohapatra:1974gc,Senjanovic:1975rk,Mohapatra:1977mj}, the SM fermions (including the right-handed neutrino) are assigned in left- or right-handed doublets, and the following Higgs sector has been assumed: one bidoublet to construct the Yukawa couplings of quarks and leptons, in addition to a left- and right-handed scalar triplets for seesaw neutrino masses. The $SU(2)_R\times U(1)_{B-L}$ symmetry is broken down to $U(1)_Y$, at a high energy scale, by the vacuum expectation value (VEV) of the neutral component of the right-handed triplet, while the VEVs of neutral components of the bidoublet and the left-handed triplet contribute in breaking the electroweak symmetry, $SU(2)_L\times
U(1)_Y$, down to $U(1)_{\text{em}}$. It was clear that the Higgs sector of this model is not minimal, with several neutral and singly and doubly charged components. Also, the left-handed triplet was introduced only to preserve LR symmetry, although its VEV must be fine-tuned to a very small value to avoid stringent constraints from the observed neutrino masses. Moreover, the Higgs triplets may induce tree level flavor violating processes that contradict the current experimental limits. Therefore, different variants of the conventional LR model have been considered~\cite{Deshpande:1990ip,Aulakh:1998nn,Maiezza:2010ic,Borah:2010zq,Nemevsek:2012iq,Ashry:2013loa}. 

Here, we consider an example of a LR model, with a Higgs sector consisting of one scalar bidoublet and a scalar right-handed doublet. In this case and in order to generate light neutrino masses, we adopt the inverse-seesaw (IS) mechanism~\cite{Mohapatra:1986aw,Mohapatra:1986bd,GonzalezGarcia:1988rw,Weiland:2013xia,Brdar:2018sbk}. As known, this mechanism requires introducing other singlet fermions that couple with right-handed neutrinos and have a small mass $[\sim\mathcal{O}(1)~\text{KeV}]$, which may be generated radiatively. The IS mechanism is quite motivated by having the TeV scale LR model that can be probed in current and future colliders, while in the conventional LR model, the GUT scale is the typical scale of breaking LR symmetry, where right-handed neutrino masses are generated. Moreover, in the limit of vanishing the above mentioned tiny mass, we will have massless left-handed neutrinos and the lepton number symmetry is restored. Thus, such a small scale can be considered, according to 't Hooft naturalness criteria~\cite{tHooft:1979rat}, as a natural scale of a global symmetry (lepton number) breaking. We also argue that in this class of models the tree-level flavor changing neutral current (FCNC) is under control. It turns out that the right-handed doublet is essentially decoupled from the two Higgs doublets, generated from the bidoublet; hence the Higgs sector of this model mimics the scenario of two the Higgs doublet model~\cite{Davidson:2005cw,Haber:2010bw,Haber:2006ue}. We show that the lightest $CP$-even Higgs boson, the SM-like Higgs boson, and the next lightest are generated from the neutral components of the bidoublet. For a wide range of the parameter space, one can show that the mass of the next lightest Higgs boson is of the order a few hundred~\text{GeVs}.

In this paper we analyze the discovery prospects of the next lightest $CP$-even neutral Higgs boson, $h'$, at the Large Hadron Collider (LHC). Our searches are performed by looking for resonant peaks in two processes, namely $h'\to hh\to b\bar{b}\gamma\gamma$ and $h'\to ZZ\to 4\ell~(\ell=e,~\mu)$. The analysis is pivoted on three benchmark points, with $m_{h'}=250~\text{GeV},~400~\text{GeV}$, and $600~\text{GeV}$, for a center-of-mass energy $\sqrt{s}=14~\text{TeV}$ and $L_{\text{int}}=300~{\text{fb}}^{-1}$, and $3000~{\text{fb}}^{-1}$, respectively. After imposing various sets of cuts to reduce backgrounds $(B)$ and improve the statistical significance $(\text{S}/\sqrt{\text{B}})$, where $S$ refers to the signal, we find that the SM-like Higgs boson pair production, with $b\bar{b}\gamma\gamma$ final states, is the most promising channel for probing our heavy Higgs boson at the LHC. The channel of the $Z$-pair production, decay to $4\ell$ is less significant as its cross section is very small for $m_{h'} \gsim 300~\text{GeV}$, and the associated background is quite large for $m_{h'} \simeq 200~\text{GeV}$. We show that to probe $h'$ through this channel, $L_{\text{int}}$ must be increased up to $L_{\text{int}}=3000~{\text{fb}}^{-1}$.

The paper is organized as follows. In Sec.~\ref{model} we introduce the LR model with IS mechanism. The Higgs sector of this model is discussed in Sec.~\ref{higgs}. We show that the SM-like Higgs, $h$, and the next lightest $h'$ are stemmed from the real part of neutral components of the bidoublet. Searches for $h'$ at the LHC are comprehensively studied in Sec.~\ref{higgssearch}. A detailed analysis for the SM Higgs pair production from $h'$, followed by the decays into $b\bar{b}\gamma\gamma$, is provided. We show that the total cross section of this process is of order $\mathcal{O}(1)$ fb. In general the signal of this channel is much smaller than the background, however by selecting an appropriate set of cuts we can probe the signal with a reasonable significance. We also analyze a possible signature through the $Z$-gauge boson pair production from $h'$ followed by the decays to $4\ell$. Our conclusions and final remarks are given in Sec.~\ref{conc}. 

\section{\label{model}Left-Right Model with inverse seesaw (LRIS)}
In this section we introduce a minimal left-right model with an inverse-seesaw (LRIS), which is based on gauge symmetry $SU(3)_C\times SU(2)_L\times SU(2)_R\times U(1)_{B-L}$.
The fermion content of this model is as the same as its counterpart in the conventional left-right models~\cite{Mohapatra:1974gc,Senjanovic:1975rk,Mohapatra:1977mj,Deshpande:1990ip,Aulakh:1998nn,Maiezza:2010ic,Borah:2010zq,Nemevsek:2012iq}. In addition, three SM singlet fermions $S_1$ with $B-L$ charge $=+2$ and three singlet fermions $S_2$ with $B-L$ charge $=-2$ are considered to implement the IS mechanism for neutrino masses. Note that we introduce pair of singlet fermions $S_{1,2}$ with opposite $B-L$ charges to keep the $U(1)_{B-L}$ anomaly free. The Higgs sector of the LRIS model consists of an $SU(2)_R$ scalar doublet $\chi_R$ to break $SU(2)_R\times U(1)_{B-L}$ to $U(1)_Y$ and a scalar bidoublet $\phi$ that breaks $SU(2)_L\times U(1)_Y$ down to $U(1)_{\text{em}}$, where the hypercharge $Y$ is defined by $Y/2=I^3_R+(B-L)/2$, and $I^3_R$ is the third component of the right isospin. The detailed quantum numbers of the fermions and Higgs bosons are presented in Table~\ref{tableparticle}.
\begin{table}[ht]
\begin{center}
\begin{tabular}{|c|c|c|c|c|c|c|c|c|} 
 \hline
 \textbf{Fields} & $Q_L=\begin{pmatrix}u_L \\d_L \\\end{pmatrix}$ & $Q_R=\begin{pmatrix}u_R \\d_R \\\end{pmatrix}$ & $L_L= \begin{pmatrix}\nu_L \\e_L \\\end{pmatrix}$ & $L_R= \begin{pmatrix}\nu_R \\e_R \\\end{pmatrix}$ & $S_1$ & $S_2$ & $\phi= \begin{pmatrix}\phi_1^0 &\phi_1^+\\\phi_2^- &\phi_2^0\\\end{pmatrix}$ & $\chi_R= \begin{pmatrix}\chi_R^+ \\\chi_R^0 \\\end{pmatrix}$\\ \hline
 $\mathbb{G_{LR}}$ & $({\bf3,2,1},\frac{1}{3})$ & $({\bf3,1,2},\frac{1}{3})$ & $({\bf1,2,1},-1)$ & $({\bf1,1,2},-1)$ & $({\bf1,1,1},-2)$ & $({\bf1,1,1},2)$ & $({\bf1,2,2},0)$ & $({\bf1,1,2},1)$ \\ 
 \hline
\end{tabular}
\caption{\label{tableparticle} The LRIS particle content and its representations and quantum numbers with respect to the LR gauge group $\mathbb{G_{LR}}=SU(3)_C\times SU(2)_L\times SU(2)_R\times U_{B-L}$.}
\end{center}
\end{table}
Also, in order to forbid a mixing mass term $M \bar{S^c_1} S_{2}$ that could spoil the IS mechanism, as discussed in~\cite{Khalil:2010iu}, a $Z_2$ discrete symmetry is used, where all particles have even charges except $S_1$, which has an odd charge. Based on the above particles and their charge assignments, the most general left-right symmetric Yukawa Lagrangian accounting for the scalar-fermion interactions and containing the IS neutrino mass terms is given by
\begin{equation}
\mathcal{L}_{\text{Y}}=\sum_{i,j=1}^3 y_{ij}^L \bar{L}_{Li}\phi L_{Rj} +\tilde{y}_{ij}^L \bar{L}_{Li} \tilde{\phi} L_{Rj} + y_{ij}^Q \bar{Q}_{Li}\phi Q_{Rj} +\tilde{y}_{ij}^Q \bar{Q}_{Li} \tilde{\phi} Q_{Rj} 
+y^s_{ij} \bar{L}_{Ri} \tilde{\chi}_R S^c_{2j}+H.c.,
\end{equation}
where $i,j=1\ldots3$ are family indices, $\widetilde{\phi}$ is the dual bidoublet of the scalar bidoublet $\phi$, defined as $\widetilde{\phi}=\tau_2\phi^*\tau_2$, and $\widetilde{\chi}_{R}$ is the dual doublet of the scalar doublet $\chi_{R}$, given by $\widetilde{\chi}_{R}=i\tau_2\chi^*_{R}$. A detailed discussion on the Higgs potential and the associated VEVs will be given in the next section. Here, we assume a nonvanishing VEV of $\chi_R$, $\langle \chi_R \rangle=v_R/\sqrt{2}$ of an order TeV to break the right-handed electroweak sector together with $B-L$. In addition, the VEVs of $\phi$, given by $\langle \phi \rangle=\text{diag}(k_1/\sqrt{2},k_2/\sqrt{2})$, are of an order $\mathcal{O}(100)~\text{GeV}$ to break the electroweak symmetry of the SM.

After $B-L$ symmetry breaking, a small mass term $\mu_s \bar{S^c_2} S_2$ (and plausibly $\mu'_s \bar{S^c_1} S_1$) is generated from a nonrenormalizable term (of dimension seven $\propto \chi_R^4 \bar{S^c_2}S_2/M^3$), which implies that $\mu_s=\lambda_s v_R^4/M^3 \lsim \mathcal{O}(1)~\text{KeV}$~\cite{Abdallah:2011ew}, where $\lambda_s$ is an interaction effective dimensionless coupling. For $v_R\sim\mathcal{O}(10^{3})~\text{GeV}$, one finds that $\mu_s\sim\mathcal{O}(10^{-7})~\text{GeV}$ if $\lambda_s/M^3\sim\mathcal{O}(10^{-19})~\text{GeV}$. Thus, if $\lambda_s\sim\mathcal{O}(1)$, the nonrenormalizable scale $M$ is an intermediate scale, namely $M\sim\mathcal{O}(10^{3})~\text{TeV}$. While for $\lambda_s\sim\mathcal{O}(10^{-3})$, one finds $M\sim\mathcal{O}(10)~\text{TeV}$. As a result, the Lagrangian of neutrino masses is given by
\be 
{\cal L}_m^\nu = M_D \bar{\nu}_L \nu_R + M_R \bar{\nu}^c_R S_2 + \mu_s \bar{S^c_2} S_2+H.c., 
\ee
where the $3\times 3$ matrix $M_D=v(y^L s_{\beta}+\tilde{y}^L c_{\beta})/\sqrt{2}$ is the Dirac neutrino mass matrix and the $3\times 3$ matrix $M_{R}=y^s v_R/\sqrt{2}$. This is the standard neutrino IS matrices~\cite{Mohapatra:1986aw,Mohapatra:1986bd,GonzalezGarcia:1988rw,Weiland:2013xia}. Here, we have fixed the VEVs $ k_{1,2}$ such that $v^2 =k_1^2 + k_2^2 $ and $k_1=vs_{\beta},~k_2=vc_{\beta}$, where we define $s_x=\sin x,~c_x=\cos x$, and $t_x=\tan x$, henceforth. In this regard, the following $9\times 9$ neutrino mass matrix can be written as $\bar{\psi}^c\mathcal{M}_\nu\psi$ with the flavor basis $\psi=(\nu_L^c,\nu_R,S_2)$ and
\begin{equation}
\mathcal{M}_\nu=
\begin{pmatrix}
0 & M_D & 0 \\
M_D^T & 0 & M_{R}\\
0 & M_{R}^T & \mu_s
\end{pmatrix}.
\end{equation}
The diagonalization of this mass matrix leads to the physical light and heavy neutrino states $\nu_{\ell_i},\nu_{h_j},~i=1\ldots3,~j=1\ldots6$, with the following mass eigenvalues:
\begin{align}
\label{nulmass}
m_{\nu_{\ell_i}}&=M_D M_{R}^{-1} \mu_s (M_{R}^T)^{-1} M_D^T,\quad i=1\ldots3,\\
\label{nuhmass}
m_{\nu_{h_j}}^2&=M_{R}^2+M_D^2,\quad j=1\ldots6.
\end{align}

On the other hand, after electroweak symmetry breaking, quarks and charged leptons acquire their masses via Higgs mechanism, as follows:
\begin{align}
\label{fermmassmat1}
M_u&=\frac{v}{\sqrt{2}}(y^Q s_{\beta}+\tilde{y}^Q c_{\beta}),\\
M_d &=\frac{v}{\sqrt{2}} (y^Q c_{\beta}+\tilde{y}^Q s_{\beta}),\\
M_{\ell} &=\frac{v}{\sqrt{2}}(y^L c_{\beta}+\tilde{y}^L s_{\beta}).
\label{fermmassmat2}
\end{align}
In contrast to the SM, fermions acquire their masses from different VEVs sources as shown in Eqs.~(\ref{fermmassmat1})-(\ref{fermmassmat2}). This allows the existence of a tree-level FCNC induced by neutral Higgs bosons exchange between different fermion families, which is severely constrained by experiments~\cite{Aad:2019pxo}. The physical fermions' masses are given after diagonalization by
\begin{equation}\label{fermmass}
M_u^{\text{diag}}=V^{u\dag}_L M'^u V^u_R,\quad M_d^{\text{diag}}=V^{d\dag}_L M'^d V^d_R,\quad M_{\ell}^{\text{diag}}=V^{\ell\dag}_L M'^\ell V^\ell_R\,.
\end{equation}
In this case, the quark Yukawa couplings can be written as 
\begin{align}
\label{yukmat}
y^Q&=-\frac{\sqrt{2}}{v c_{2\beta}}( s_{\beta}V^u_L M^u V^{u\dag}_R- c_{\beta}V^d_L M^d V^{d\dag}_R),\\
\tilde{y}^Q&=-\frac{\sqrt{2}}{ v c_{2\beta}}( c_{\beta}V^u_L M^u V^{u\dag}_R- s_{\beta}V^d_L M^d V^{d\dag}_R).
\end{align}
These Yukawa couplings lead to the following interactions between quarks and neutral Higgs bosons
\begin{align}
{\cal L}_{\text{int}} &= \frac{-\sqrt{2}}{v c_{2\beta}} \left[ \bar{u}_L (M_u^{\text{diag}} ( s_{\beta} \phi_1^0 - c_{\beta} \phi_2^{0*}) + V^{L}_{\text{CKM}} M_d^{\text{diag}} V^{R}_{\text{CKM}}(- c_{\beta} \phi_1^0 + s_{\beta} \phi_2^{0*})) u_R \right]\,\nonumber\\
&-\frac{\sqrt{2}}{v c_{2\beta}} \left[ \bar{d}_L (M_d^{\text{diag}} ( s_{\beta} \phi_1^0 - c_{\beta} \phi_2^{0*}) + V^{L}_{\text{CKM}} M_u^{\text{diag}} V^{R}_{\text{CKM}} (- c_{\beta} \phi_1^0 + s_{\beta} \phi_2^{0*})) d_R \right],
\end{align}
where $V_{\text{CKM}}^{L,R}=V^{u\dag}_{L,R} V^d_{L,R}$ are the mixing matrices for left and right quarks, respectively. It is clearly that the second nondiagonal terms in the above equations are sources of FCNC, which could be quite dangerous if one considers $V_{\text{CKM}}^R$ with large off diagonal entries. 
We assume that our LR model is derived from a GUT model, like $SO(10)$; therefore a universal gauge coupling is naturally adopted: $g_L=g_R$. Furthermore, we consider a truly minimal Higgs sector made up of one bidoublet $\phi$ and one RH doublet $\chi_R$. As shown in~\cite{Senjanovic:2014pva,Maiezza:2020fcv}, in this class of models with $g_L=g_R$, the RH-mixing matrix $V^R_{\text{CKM}}$ can be calculated in terms of the LH-mixing matrix $V^L_{\text{CKM}}$ and $V^R_{\text{CKM}} \simeq V^L_{\text{CKM}}$, at the leading order. 
As a result, the flavor violation is under control and no longer dangerous, because it will be proportional to some power of the Wolfenstein parameter. $\lambda \sim 0.12$~\cite{PhysRevLett.51.1945,Tanabashi:2018oca}. 
For example, the $b-s$ FCNC transition via a neutral Higgs boson $H_i~(i=1\ldots3)$ will be proportional to $(V^{L}_{\text{CKM}})_{3i} (M_u^{\text{diag}})_{ii} (V^{R}_{\text{CKM}})_{i2} =(V^{L}_{\text{CKM}})_{3i} (M_u^{\text{diag}})_{ii} (V^{L}_{\text{CKM}})_{i2}$. Thus, with the neutral Higgs mixing $Z^H_{ij} < 1$, one finds that $\frac{1}{ vc_{2\beta}}(-c_{\beta} Z^H_{i1}+s_{\beta} Z^H_{i2})(m_u \lambda^4 + m_c \lambda^2 + m_t \lambda^2)\ll \lambda^2$, which is small enough and does not generate any conflict with recent experimental observations~\cite{Aad:2019pxo,Park:2018onm}.

Now we turn to the gauge sector. The scalar bidoublet $\phi$ mixes the left and right gauge bosons. Thus, one can show that the symmetric mass matrix for neutral left, right and $B-L$ gauge bosons basis $(W_{R\mu}^3,V_\mu,W_{L\mu}^3)$, is given by
\begin{equation}
M_{ZZ'}^2=\frac{1}{4}
\begin{pmatrix}
 g_R^2(v_R^2+v^2) & -g_{BL} g_R v_R^2 & -g_L g_R v^2 \\
 . & g_{BL}^2 v_R^2 & 0 \\
 . & . & g_L^2v^2
\end{pmatrix}.
\end{equation}
Notice that we exploit the symmetry of the matrix to write dots for the lower off diagonal elements instead of repeating them through the text. After the first stage of left-right symmetry breaking down to the SM symmetry, both the right and $B-L$ gauge bosons $W_{R\mu}^3$ and $V_\mu$ mix with an angle $\varphi$ whose $s_\varphi=g_{BL}/\sqrt{g_R^2+g_{BL}^2}$ to give the massless hypercharge $U(1)_Y$ gauge boson $B_\mu$, while the right neutral gauge boson $Z_{R\mu}$ becomes massive. Then, the SM gauge symmetry is broken and $W^3_{L\mu}$, and $B_\mu$ mix with the Weinberg angle $\sin\theta_w \equiv s_w=e/g_L$ and give the photon $A_\mu$ and the left weak boson $Z_{L\mu}$. However, both $Z_{L\mu}$ and $Z_{R\mu}$ mix by the scalar bidoublet $\phi$, as mentioned above, and give the physical states of $Z_\mu$ and $Z'_\mu$ bosons as follows
\begin{equation}\label{zlzramix}
\begin{pmatrix}Z_{R\mu} \\ B_\mu\end{pmatrix}=
\begin{pmatrix}c_\varphi & -s_\varphi \\ s_\varphi & c_\varphi\end{pmatrix} 
\begin{pmatrix}W^3_{R\mu} \\ V_\mu\end{pmatrix},\quad
\begin{pmatrix}Z_{L\mu} \\ A_\mu\end{pmatrix}=
\begin{pmatrix}c_w & -s_w \\ s_w & c_w\end{pmatrix} 
\begin{pmatrix}W^3_{L\mu} \\ B_\mu\end{pmatrix},\quad
\begin{pmatrix}Z_\mu\\Z'_\mu\end{pmatrix}=
\begin{pmatrix}c_\vartheta & -s_\vartheta \\s_\vartheta & c_\vartheta\end{pmatrix} 
\begin{pmatrix}Z_{L\mu}\\Z_{R\mu}\end{pmatrix}.
\end{equation}
This leads to the total mixing $(W^3_{R\mu},V_\mu,W^3_{L\mu})^T=Z^Z(Z'_\mu,Z_\mu,A_\mu)^T$, where $Z^{ZT} M_Z^2 Z^Z=\text{diag}(M_{Z'}^2,M_Z^2,0)$ and the rotation matrix $Z^Z$ is
\begin{equation}\label{zzpamix}
Z^Z=
\begin{pmatrix}
c_\vartheta c_\varphi-s_\vartheta s_w s_\varphi & -s_\vartheta c_\varphi-c_\vartheta s_w s_\varphi & c_w s_\varphi \\
-c_\vartheta s_\varphi-s_\vartheta s_w c_\varphi & s_\vartheta s_\varphi-c_\vartheta s_w c_\varphi & c_w c_\varphi \\
s_\vartheta c_w & c_\vartheta c_w & s_w
\end{pmatrix},
\end{equation}
where the $Z-Z'$ mixing angle $\vartheta$ is given by
\begin{equation}
\tan2\vartheta=\frac{2M_{LR}}{M_{LL}-M_{RR}}=\frac{2g_{2}^3 \sqrt{g_{2}^2+2g_{BL}^2}}{(g_{2}^2+g_{BL}^2)^2 (\frac{v_R}{v})^2-2g_{2}^2 g_{BL}^2},
\end{equation}
and the $(Z_{L\mu},Z_{R\mu})$ symmetric mass matrix elements are
\begin{align}
M_{LL}&=\frac{ g_{2}^2 v^2 (g_{2}^2+2g_{BL}^2)}{4(g_{2}^2+g_{BL}^2)},\\
M_{LR}&=M_{RL}=\frac{-g_{2}^3 v^2 \sqrt{g_{2}^2+2g_{BL}^2}}{4(g_{2}^2+g_{BL}^2)},\\
M_{RR}&=\frac{ g_{2}^4 v^2+(g_{2}^2+g_{BL}^2)^2 v_R^2}{4(g_{2}^2+g_{BL}^2)}.
\end{align}
The mass eigenvalues are given by 
\begin{equation}
M_{Z,Z'}^2=\frac{1}{2}(M_{LL}+M_{RR}\mp (M_{RR}-M_{LL})\sqrt{1+\tan^22\vartheta}).
\end{equation}
The LHC search for the $Z'$ gauge boson is rather model dependent; in our case, we found that the following limits are imposed on $M_{Z'}$: $0.8\lsim M_{Z'}\lsim 4.5~\text{TeV}$~\cite{Sirunyan:2021bzu,Aaboud:2017cxo}.

The charged gauge boson symmetric mass matrix in basis $(W_{L\mu}^{\pm},W_{R\mu}^{\pm})$ is given by
\begin{equation}
M_{WW'}^2=\frac{1}{4}
\begin{pmatrix}
 g_L^2v^2 & -g_L g_R v^2s_{2\beta}\\
 . & g_R^2(v^2+v_R^2)
\end{pmatrix}.
\end{equation}
Thus, the $W-W'$ diagonalization mixing angle $\xi$ is
\begin{equation}
\tan 2\xi=\frac{2 g_L g_Rs_{2\beta}}{g_R^2(1+(\frac{v_R}{v})^2)-g_L^2},\end{equation}
and the physical gauge bosons masses $M_{W,W'}$ are approximately
\begin{equation}
M_{W,W'}^2=\frac{1}{8}(g_L^2v^2+g_R^2(v^2+v_R^2)\mp (g_R^2(v^2+v_R^2)-g_L^2v^2)\sqrt{1+\tan^22\xi}).
\end{equation}
or approximately with $g_L=g_R$,
\begin{equation}
M_{W}=\frac{g_2 v}{2},\quad M_{W'}=\frac{g_2\sqrt{v^2+v_R^2}}{2}.
\end{equation}

The LHC's direct searches for $W'$ impose stringent constraints on $M_{W'}$. These constraints, however, are model dependent, as they are determined by the assumptions imposed on the gauge couplings and the dominated channels of $W'$ decays. In the LR model, the decay channel of $W'$ to $\nu_R$ may be dominant, suppressing other decay channels to SM leptons or quarks. As a result, the bounds on $M_{W'}$ are relaxed, as highlighted in Ref.~\cite{Frank:2018ifw}. In this case, a conservative bound on the mass of the gauge boson $W'$ is of order $2~\text{TeV}$.

\section{\label{higgs}Higgs Sector in the LRIS}
As mentioned in the previous section, the Higgs sector of the LRIS consists of a bidoublet $\phi$ and a right doublet $\chi_R$ as in Table~\ref{tableparticle}, and the gauge symmetries $SU(2)_R\times U(1)_{B-L}$ are spontaneously broken to $U(1)_Y$ through the VEV of $\chi_R$, then the $SU(2)_L\times U(1)_Y$ symmetries are broken by VEVs of $\phi$. The most general Higgs potential that is invariant under the above mentioned symmetries (gauge and discrete) is given by~\cite{Borah:2010zq}
\begin{align}\label{scalarpot}
V(\phi, \chi_R) \nonumber &= \mu_1 \text{Tr}(\phi^{\dagger} \phi) +\mu_2 [\text{Tr}(\tilde{\phi} \phi^{\dagger})+\text{Tr}(\tilde{\phi}^{\dagger} \phi)] + \lambda_1(\text{Tr}(\phi^{\dagger} \phi))^2+ \lambda_2[(\text{Tr}(\tilde{\phi}\phi^\dagger))^2+(\text{Tr}(\tilde{\phi}^\dagger\phi))^2]\\ 
&+ \lambda_3 \text{Tr}(\tilde{\phi}\phi^{\dagger}) \text{Tr}(\tilde{\phi}^{\dagger}\phi) +\lambda_4 \text{Tr}(\phi \phi^{\dagger}) (\text{Tr}(\tilde{\phi}\phi^{\dagger})+\text{Tr}(\tilde{\phi}^{\dagger}\phi))+\mu_3 (\chi_R^{\dag} \chi_R) +\rho_1(\chi_R^{\dag} \chi_R)^2 \nonumber \\
& + \alpha_1 \text{Tr}(\phi^{\dagger} \phi) (\chi_R^{\dag} \chi_R)+ \alpha_2 (\chi_R^{\dagger} \phi^{\dagger} \phi \chi_R)+\alpha_3 (\chi_R^{\dagger}\tilde{\phi}^{\dagger} \tilde{\phi} \chi_R)+\alpha_4 (\chi_R^{\dagger} \phi^{\dagger} \tilde{\phi} \chi_R +H.c.).
\end{align}
It is worth mentioning here that the potential parameters in (\ref{scalarpot}) are constrained by the spectrum and minimization and boundedness from below conditions of the potential provided in Appendixes~\ref{apptdplmin}~and~\ref{appcopbd}.
\subsection{Higgs masses and mixing}
It is worth noting that before symmetry breaking there were $12$ scalar degrees of freedom: $8$ of $\phi$ and $4$ of $\chi_R$. After symmetry breaking, two neutral components and four charged components of these degrees of freedom are eaten by the neutral gauge bosons: $Z_\mu$ and $Z'_\mu$ and the charged gauge bosons: $W^\pm_\mu$ and $W'^\pm_\mu$ to acquire their masses, respectively. Therefore in this class of models, six scalars remain as physical Higgs bosons; as we show, two of them are charged Higgs bosons, one is a pseudoscalar Higgs boson, and the remaining three give $CP$-even neutral Higgs bosons.
\begin{enumerate}
\bigskip

\item \textbf{Charged Higgs bosons}

It can be easily seen that the symmetric mass matrix of the charged Higgs bosons in the basis $(\phi_1^{\pm},\phi_2^{\pm},\chi_R^{\pm})$ is given by 
\begin{equation}\label{mhpm2}
M_{H^\pm}^2=\frac{\alpha_{32}}{2}\begin{pmatrix}
\frac{v_R^2 s_{\beta}^2}{c_{2\beta}} & \frac{v_R^2 s_{2\beta}}{2c_{2\beta}} & -v v_R s_{\beta} \\
 . & \frac{v_R^2 c_{\beta}^2}{c_{2\beta}} & -v v_R c_{\beta} \\
 . & . & v^2 c_{2\beta}
\end{pmatrix}.
\end{equation}
Notice that since $s_\beta\ll1$, the entries of the above matrix are of following orders: $(M_{H^\pm}^2)_{11}\ll v_R^2,~(M_{H^\pm}^2)_{22}\approx v_R^2$ and $(M_{H^\pm}^2)_{33}\approx v^2$, while the off diagonal elements are given by $(M_{H^\pm}^2)_{12}\ll v_R^2,~(M_{H^\pm}^2)_{13}\lsim vv_R$ and $(M_{H^\pm}^2)_{23}\approx vv_R$. The matrix $M_{H^\pm}^2$ can be diagonalized by a unitary matrix,
\begin{equation}
Z^{H^\pm}=\begin{pmatrix}
\frac{v c_{2\beta}}{\sqrt{v^2c_{2\beta}^2+v_R^2s_{\beta}^2}} & 0 & \frac{v_R s_{\beta}}{\sqrt{v^2c_{2\beta}^2+v_R^2s_{\beta}^2}} \\
-\frac{\frac{1}{2}v_R^2 s_{2\beta}}{\sqrt{(v^2c_{2\beta}^2+v_R^2s_{\beta}^2)(v^2c_{2\beta}^2+v_R^2)}} 
& \sqrt{\frac{v^2c_{2\beta}^2+v_R^2s_{\beta}^2}{v^2c_{2\beta}^2+v_R^2}}
& \frac{v v_R c_{\beta}c_{2\beta}}{\sqrt{(v^2c_{2\beta}^2+v_R^2s_{\beta}^2)(v^2c_{2\beta}^2+v_R^2)}}
\\
-\frac{v_R s_{\beta}}{\sqrt{v^2c_{2\beta}^2+v_R^2}} & -\frac{v_R c_{\beta}}{\sqrt{v^2c_{2\beta}^2+v_R^2}} 
& \frac{v c_{2\beta}}{\sqrt{v^2c_{2\beta}^2+v_R^2}} 
\end{pmatrix},
\end{equation}
such that $Z^{H^\pm}M_{H^\pm}^2Z^{H^\pm T}=\text{diag}(0,0,m_{H^\pm}^2)$ and $(\phi_1^{\pm},\phi_2^{\pm},\chi_R^{\pm})^T=Z^{H^\pm T}(G_1^{\pm},G_2^{\pm},H^{\pm})^T$, where $G_1^{\pm}$ and $G_2^{\pm}$ are the charged Goldstone bosons eaten by the charged gauge bosons $W_\mu$ and $W'_\mu$ to acquire their masses via the Higgs mechanism. In addition, it remains one massive eigenstate of a physical charged Higgs boson $H^{\pm}$, with the following mass:
\begin{equation}
m_{H^{\pm}}^2 = \frac{\alpha_{32}}{2}\left(\frac{ v_R^2}{c_{2\beta}}+v^2c_{2\beta}\right),
\end{equation}
where $\alpha_{32}=\alpha_3-\alpha_2$. Thus, if $v_R\sim\mathcal{O}(\text{TeV})$ and $\alpha_{32}\sim\mathcal{O}(10^{-2})$, then the charged Higgs boson mass is on the order of hundreds~\text{GeV} as in Table~\ref{Bnchmrkpnts}. 
The physical charged Higgs boson is given through the rotation matrix $Z^{H^\pm}$ by 
\begin{equation}
H^\pm=Z^{H^\pm}_{13}\phi_1^\pm+Z^{H^\pm}_{23}\phi_2^\pm+Z^{H^\pm}_{33}\chi_R^\pm,
\end{equation}
where $Z^{H^\pm}_{13}\approx1$, while $Z^{H^\pm}_{23}\approx v/(v_R t_\beta) < 1$ and $Z^{H^\pm}_{33}\approx v/v_R \ll 1$. Therefore, the charged Higgs boson is mainly made of $\phi_1^\pm$.

\item \textbf{$CP$-odd Higgs bosons}

Now, we consider the neutral scalar fields and their masses. This can be obtained if one fluctuates the neutral components of the bidoublet $\phi$ and the doublet $\chi_{R}$ around their vacua as follows:
\be
\phi_i^0 = \frac{1}{\sqrt{2}} ( v_i +\phi_i^{0R} +i \phi_i^{0I}),
\ee
where $\phi_i=\phi_{1,2},\chi_{R}$ and $v_i=k_{1,2},v_{R}$. In this case, the symmetric mass matrix of the $CP$-odd Higgs bosons is given by
\begin{equation}
(M_{A}^2)_{ij} = \frac{\partial^2 V(\phi,\chi_{L,R})}{\partial \phi_i^{0I} \partial \phi_j^{0I}} \Big\vert_{\langle \phi_{i,j}^{0R}\rangle = \langle \phi_{i,j}^{0I}\rangle =0}.
\end{equation}
Therefore, in the basis $(\phi_1^{0I}, \phi_2^{0I}, \chi_R^{0I})$, the pseudoscalar Higgs bosons mass matrix is given by 
\begin{equation}
M_{A}^2=\frac{1}{2}\left(\frac{v_R^2 \alpha_{32}}{c_{2\beta}}-4v^2(2\lambda_2-\lambda_3)\right)
\begin{pmatrix}
c_{\beta}^2 & s_\beta c_\beta & 0 \\
 . & s_{\beta}^2 & 0 \\
 . & . & 0 
\end{pmatrix}, 
\end{equation}
which can be diagonlized by a unitary matrix,
\begin{equation}
Z^A=\begin{pmatrix}
 0 & 0 & 1 \\
 -s_{\beta} & c_{\beta} & 0 \\
 c_{\beta} & s_{\beta} & 0
\end{pmatrix},
\end{equation}
such that $Z^A M_A^2 Z^{AT}=\text{diag}(0,0,m_A^2)$ and $(\phi_1^{0I},\phi_2^{0I},\chi_R^{0I})^T=Z^{AT}(G_1^0,G_2^0,A)^T$, where $G_1^0$ and $G_2^0$ are the neutral Goldstone bosons eaten by the two neutral gauge bosons $Z_\mu$ and $Z'_\mu$ to acquire their masses, and the following nonzero eigenvalue is the physical mass of the pseudoscalar boson $A$,
\begin{equation}
m_{A}^2 =\frac{1}{2}\left(\frac{v_R^2}{c_{2\beta}}\alpha_{32}-4v^2(2\lambda_2-\lambda_3)\right).
\end{equation}
Thus, for $v_R\sim\mathcal{O}(\text{TeV})$ and $\alpha_{32} \sim\mathcal{O}(10^{-2})$, the pseudoscalar Higgs boson mass is on the order of a few hundred~\text{GeVs}, as well as in Table~\ref{Bnchmrkpnts}. 
The physical pseudoscalar Higgs boson is given through the rotation matrix $Z^A$ by 
\begin{equation}
A=c_\beta\phi_1^{0I}+s_\beta\phi_2^{0I}.
\end{equation}
According to our choice $s_\beta\ll1$ and $A\approx \phi_1^{0I}.$

\item \textbf{$CP$-even Higgs bosons}

Finally, we consider the $CP$-even Higgs bosons. Similar to the $CP$-odd Higgs bosons, the $(3\times 3)$ symmetric mass matrix of $CP$-even Higgs bosons is given by
\be\label{cpevenhgsms}
(M_{H}^2)_{ij} = \frac{\partial^2 V(\phi,\chi_{L,R})}{\partial \phi_i^{0R} \partial \phi_j^{0R}} \Big\vert_{\langle \phi_{i,j}^{0R}\rangle = \langle \phi_{i,j}^{0I}\rangle =0},
\ee
with the elements
\begin{align}
m_{11}&= 2v^2 (\lambda_{1}s_\beta^2+\lambda_{23}c_\beta^2+\lambda_{4} s_{2\beta})+\frac{1}{4}(\frac{1}{c_{2\beta}}+1)\alpha_{32} v_R^2, \\
m_{12}&=m_{21}= v^2 ((\lambda_{1}+\lambda_{23})s_{2\beta}+2\lambda_{4})-\frac{1}{4} \alpha_{32} v_R^2 t_{2\beta}, \\
m_{13}&=m_{31}= v v_R (\alpha_{13} s_{\beta}+\alpha_4 c_{\beta}), \\
m_{22}&= 2v^2 (\lambda_{1}c_\beta^2+\lambda_{23}s_\beta^2+\lambda_{4} s_{2\beta})+\frac{1}{4}(\frac{1}{c_{2\beta}}-1)\alpha_{32} v_R^2, \\
m_{23}&=m_{32}= v v_R (\alpha_{12} c_{\beta}+\alpha_4 s_{\beta}), \\
m_{33}&= 2 \rho_1 v_R^2,
\end{align}
where $\alpha_{1i}=\alpha_1+\alpha_i,~i=2,3$ and $\lambda_{23}=2\lambda_2+\lambda_3$.
This matrix can be diagonalized by a unitary transformation matrix $Z^{H}$ such that $Z^H M_H^2 Z^{HT}=\text{diag}(m_{H_1}^2,m_{H_2}^2,m_{H_3}^2)$ and $(\phi_1^{0R},\phi_2^{0R},\chi_R^{0R})^T=Z^{HT}(H_1,H_2,H_3)^T$. The coefficients of the rotation matrix $Z^H$ are given explicitly in Appendix~\ref{appdet}. After rotation, we obtain three massive neutral Higgs bosons
\begin{equation}\label{hmix}
H_i = Z^H_{ij} \phi_j^{0R}.
\end{equation}
Here, a few remarks are in order (for $t_\beta\ll1$): $(i)$ To obtain the SM-like Higgs boson $h$ and another Higgs boson $h'$ of order of a few~\text{GeVs} we take $\lambda_1\sim\mathcal{O}(10^{-1})$ and $\alpha_{32}\sim\mathcal{O}(10^{-2})$ so that $m_{11},m_{22}\sim\mathcal{O}(\text{GeV}^2)$. 
$(ii)$ Taking $\rho_1\sim\mathcal{O}(10^{-1})$ alongside $\alpha_{32}\sim\mathcal{O}(10^{-2})$ leads to small mixing between $\chi_R^{0R}$ and the bidoublet components $\phi_{1,2}^{0R}$ such that $m_{13}^2/(m_{11}m_{33})\ll1$ and $m_{23}^2/(m_{22}m_{33})\ll1$. This leads to the small mixing of $\chi_R^{0R}$ and the light states $h,h'$ and small mixing between the heaviest state $H_3$ and the bidoublet components $\phi_{1,2}^{0R}$, \ie~$Z^H_{13},Z^H_{23},Z^H_{31},Z^H_{32}\sim\mathcal{O}(10^{-2}-10^{-1})\ll1$, as in Table~\ref{Bnchmrkpntsmxzh}. Thus, the heaviest Higgs boson $H_3\sim\chi_R^{0R}$ with a mass $m_{H_3}^2\approx m_{33}$. 
$(iii)$ Moreover, if we assume that $\lambda_{23}\geq\lambda_1$, one finds that $m_{11}>m_{22}$ and $m_{12}^2/(m_{11}m_{22})\ll1$, which implies that the mixing $Z^H_{11},Z^H_{22}\sim\mathcal{O}(10^{-2}-10^{-1})<1$ as in Table~\ref{Bnchmrkpntsmxzh}. Accordingly, the SM-like Higgs boson can be defined as $H_1\sim\phi_2^{0R}$ and the next lightest Higgs boson is given by $H_2\sim\phi_1^{0R}$, with masses $m_{H_1}^2\approx m_{22}-m_{12}^2/(m_{11}-m_{22})<m_{H_2}^2\approx m_{11}+m_{12}^2/(m_{11}-m_{22})$. In addition, this keeps the SM-like Higgs couplings with the SM fermions and gauge bosons intact. 
$(iv)$ For small values of $\lambda_4$, one finds that $H_1$ could be the lightest while $H_2$ may represent the SM-like Higgs boson. However, for most of the parameter space ($\lambda_4$, $\alpha_{32}$, $t_\beta$) we have $H_1$ as the SM-like Higgs boson, and this is what we adopted here as in Eq.~(\ref{fixpar}) below. 
$(v)$ Due to the mass hierarchy considered here, values of $\alpha_{12,13}$ and $\alpha_4$ are not constrained for the mass spectrum. But in our LHC analysis below, we see that large values of $\alpha_4$ and $\lambda_4$, as in Eq.~(\ref{fixpar}) and Table~\ref{Bnchmrkpnts}, are preferable for large coupling $g_{h'hh}$ in Eq.~(\ref{hphh}), and hence for large $\sigma(h'\to hh)$. Lastly, we should mention here that all parameters are constrained by the copositivity conditions of Appendix~\ref{appcopbd}.

As emphasized, the lightest eigenstate $H_1 \equiv h$ will be the SM-like Higgs boson that we fix its mass with $m_h=125~\text{GeV}$~\cite{Aad:2012tfa,Chatrchyan:2012ufa}. This condition can be used to fix the value of one of the involved parameters, $\lambda_1$. The other two eigenvalues are given by 
\begin{equation}
m_{H_{2,3}}^2=\frac{1}{2}\left(T^h-m_h^2\mp\sqrt{(T^h-m_h^2)^2-\frac{4D^h}{m_h^2}}\right),
\label{mh23}
\end{equation}
where the trace $T^h$ of the Higgs matrix $M_H^2$ is
\begin{align}
T^h=\text{Tr}(M_H^2)&=2v^2 (\lambda_1+\lambda_{23}+2\lambda_4 s_{2\beta})+(\frac{\alpha_{32}}{2c_{2\beta}}+2\rho_1) v_R^2 ,
\end{align}
and $D^h$ is the determinant of $M_H^2$, which is explicitly given in Appendix~\ref{appdet}. 
From Eq.~(\ref{mh23}), one can show that the next lightest $CP$-even neutral Higgs boson, $H_2 \equiv h'$, could have a mass of order a few hundred~\text{GeVs}, as shown in Fig.~\ref{massvspar}~(left). In this figure, $m_{h'}$ is given as function of $\alpha_{32}$, which is one of the relevant parameters in the scalar potential, with choosing $\lambda_{23}\in[-0.1,3]$ and varying other parameters in the following ranges upon our discussion above:
\begin{equation}\label{fixpar}
\lambda_1\in[0.18,0.30],\quad \lambda_4\in[0.70,0.99],\quad \alpha_1\in[0.06,0.16],\quad \alpha_4\in[0.60,0.99],\quad \rho_1 \in[0.08,0.14]~.
\end{equation}
Moreover, a huge scan over the parameter space of the LRIS was conducted taking into account the recent LHC contraints using integratively \higgsbounds~and \higgssignals~programs as explained below~\cite{Bechtle:2013wla,Bechtle:2013xfa}. The scan confirmed our previous discussions and ranges considered. We chose the three benchmark points in Table~\ref{Bnchmrkpnts} with optimized cross sections in both signals of interest in our analysis below. We also considered our $h'$ in the three mass ranges $m_{h'}=250,400~\text{GeV}$ and $600~\text{GeV}$. In Fig.~\ref{massvspar}~(left), we circled the three benchmark points of Table~\ref{Bnchmrkpnts}.
\begin{figure}[ht]
\centering
\includegraphics[scale=.85]{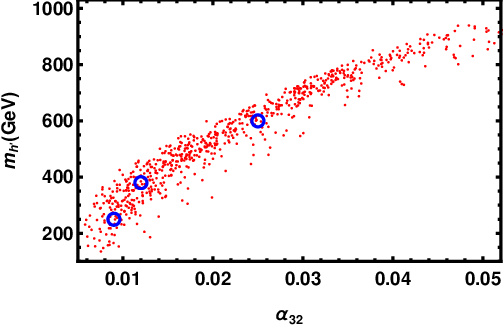}~\quad~\includegraphics[scale=.63]{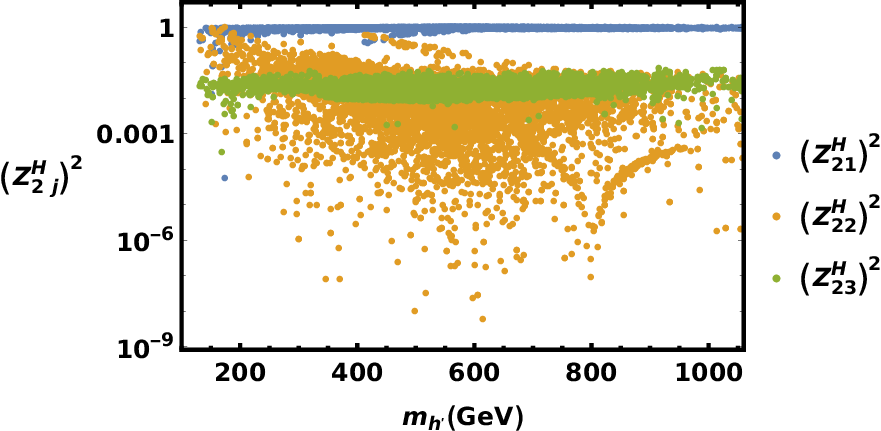}
\caption{\label{massvspar}Left: The next lightest Higgs boson mass $m_{h'}$ as a function of the most relevant parameter $\alpha_{32}$. The three benchmark points under consideration of Table~\ref{Bnchmrkpnts} are surrounded by blue circles.~Right: $h'$ mixing versus its mass $m_{h'}$. The other involved parameters are varied as in Eq.~(\ref{fixpar}).}
\end{figure}

The interactions of $h'$ with the SM fermions and gauge bosons, which are quite relevant for its search at the LHC, are given in terms of the mixing coupling $Z^{H}_{2i}$. As explained above, the physical eigenstate $h'$ is given by the superposition of real parts of neutral components of scalar doublets as follows:
\begin{equation}
h'=Z^{H}_{21} ~\phi^{0R}_1 + Z^{H}_{22} ~\phi^{0R}_2 + Z^{H}_{23} ~\chi^{0R}_R.
\end{equation}
In Fig.~\ref{massvspar}~(right), we display the mixing $Z^H_{2i}$ versus $m_{h'}$ for the same set of parameters considered in Fig.~\ref{massvspar}~(left). As can be seen from this plot, $h'$ is essentially generated from $\phi_1$ with smaller contributions from the real components of $\phi_2$ and $\chi_R$.
\end{enumerate}
%
Before closing this section, we highlight the relevant interaction couplings of $h'$ with $Z_\mu$ gauge boson, $g_{h'ZZ}$, and with the SM-like Higgs boson, $g_{h'hh}$. These interactions are generated from kinetic terms and the scalar potential terms, respectively, and they are dominantly given by
\begin{align}
\label{hphh}
g_{h'hh} &\approx -2i Z^H_{21} Z^H_{12} \{v ((\lambda_1-\lambda_{23}) c_{\beta}+3\lambda_4 s_{\beta}) Z^H_{12} + \alpha_4 v_R Z^H_{13}\},\\
\label{hpzz}
g_{h'ZZ} &\approx \frac{i}{2} g_2^2 v (c_{\beta} Z^H_{21}+s_{\beta} Z^H_{22} ) (Z^Z_{32}-Z^Z_{12})^2,
\end{align}
where $Z^Z$ is the neutral gauge bosons mixing matrix, given in Eq.~(\ref{zzpamix}). We fix the gauge couplings as follows: $g_L=g_R=g_2=0.663,~g_{BL}=0.422$, the electroweak VEV as $v=246~\text{GeV}$, and the Weinberg angle as $s_w^2=0.230$~\cite{Tanabashi:2018oca}. In addition, the scale of LR symmetry breaking is fixed by $v_R=6400~\text{GeV}$ as given in Table~\ref{Bnchmrkpnts}. Finally, the scalar potential parameters $\alpha_i$, $\lambda_i$ and $\rho_1$ are varied within the above mentioned ranges in Eq.~(\ref{fixpar}). Also, one can show that the coupling $g_{h'ZZ}$ is typically smaller than the coupling $g_{h'hh}$. Therefore, probing $h'$ through its $ZZ$-decay channel may not be promising, as it will be illustrated in the next section. In Table~\ref{Bnchmrkpnts}, we fix the three benchmark points and their corresponding Higgs spectrum which are used in the LHC simulation analysis in Sec.~\ref{higgssearch}, while Table~\ref{Bnchmrkpntsmxzh} exhibits their corresponding neutral $CP$-even Higgs mixing.

\begin{table}[ht]
\begin{center}
\begin{tabular}{|c|c|c|c|c|c|c|c|c|c|c|c|c|c|c|c|c|} 
\hline
\textbf{Parameter} & $\alpha_1$ & $\alpha_2$ & $\alpha_3$ & $\alpha_4$ & $\lambda_1$ & $\lambda_2$ & $\lambda_3$ & $\lambda_4$ & $\rho_1$ & $t_{\beta}$ & \begin{tabular}{cc}$v_R$\\\text{(GeV)}\\\end{tabular} & \begin{tabular}{cc}$m_{H^{\pm}}$\\\text{(GeV)}\\\end{tabular} & \begin{tabular}{cc}$m_{A}$\\\text{(GeV)}\\\end{tabular} & \begin{tabular}{cc}$m_{h'}$\\\text{(GeV)}\\\end{tabular} & \begin{tabular}{cc}$m_{H_3}$\\\text{(GeV)}\\\end{tabular}\\\hline
\textbf{BP1} & 0.133 & 0.155 & 0.164 & 0.833 & 0.215 & 0.316 & $-0.155$ & 0.997 & 0.104 & 0.134 & 6400 & 440 & 315 & 250 & 3000 \\
\textbf{BP2} & 0.229 & 0.090 & 0.102 & 0.620 & 0.198 & 0.230 & $-0.104$ & 0.917 & 0.117 & 0.159 & 6400 & 430 & 350 & 400 & 3100 \\
\textbf{BP3} & 0.118 & 0.168 & 0.193 & 0.957 & 0.228 & 0.309 & $-0.116$ & 0.985 & 0.138 & 0.055 & 6400 & 700 & 650 & 600 & 3400 \\
\hline
\end{tabular}
\caption{\label{Bnchmrkpnts}Benchmark points (BP) and corresponding Higgs spectrum used in figures and analysis of Sec.~\ref{higgssearch}.}
\end{center}
\end{table}

\begin{table}[ht]
\begin{center}
\begin{tabular}{|c|c|c|c|c|c|c|c|c|c|} 
\hline
\textbf{Mixing} & $Z^H_{11}$ & $Z^H_{12}$ & $Z^H_{13}$ & $Z^H_{21}$ & $Z^H_{22}$ & $Z^H_{23}$ & $Z^H_{31}$ & $Z^H_{32}$ & $Z^H_{33}$ \\\hline
\textbf{BP1} & $-0.135$ & 0.989 & $-0.051$ & 0.978 & 0.125 & $-0.166$ & 0.158 & 0.072 & 0.985 \\
\textbf{BP2} & $-0.155$ & 0.987 & $-0.051$ & 0.982 & 0.148 & $-0.119$ & 0.110 & 0.068 & 0.992 \\
\textbf{BP3} & $-0.065$ & 0.997 & $-0.038$ & 0.988 & 0.059 & $-0.141$ & 0.138 & 0.047 & 0.989 \\
\hline
\end{tabular}
\caption{\label{Bnchmrkpntsmxzh}Neutral Higgs mixing corresponding to the three BPs in Table~\ref{Bnchmrkpnts}.}
\end{center}
\end{table}
It is noticeable that there are three effective parameters: $\alpha_{12},~\alpha_{13}$, and $\alpha_{23}$ involved in the above expressions. They are given in terms of the three parameters: $\alpha_{1},~\alpha_{2}$, and $\alpha_{3}$, so there is not any redundancy. Also, the effective parameter $\lambda_{23}=2\lambda_{2}+\lambda_{3}$ and $2\lambda_{2}-\lambda_{3}$ (in the pseudoscalar mass) in terms of the two parameters $\lambda_{2}$ and $\lambda_{3}$, and hence we have two independent parameters and there is no redundancy again. Finally, to satisfy the $W'$ and $Z'$ mass constraints, we should take $v_R\sim\mathcal{\text{TeV}}$.

\section{\label{higgssearch}Search for Heavy Higgs Bosons at the LHC}
\begin{figure}[ht]
\centering
\includegraphics[scale=.95]{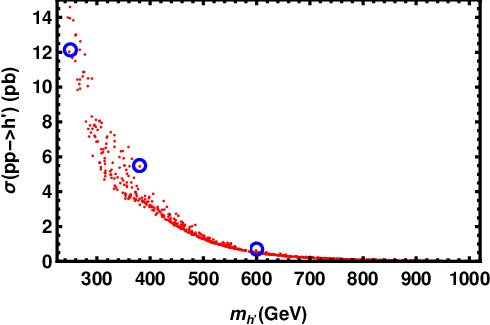}~\quad\quad~\includegraphics[scale=0.65]{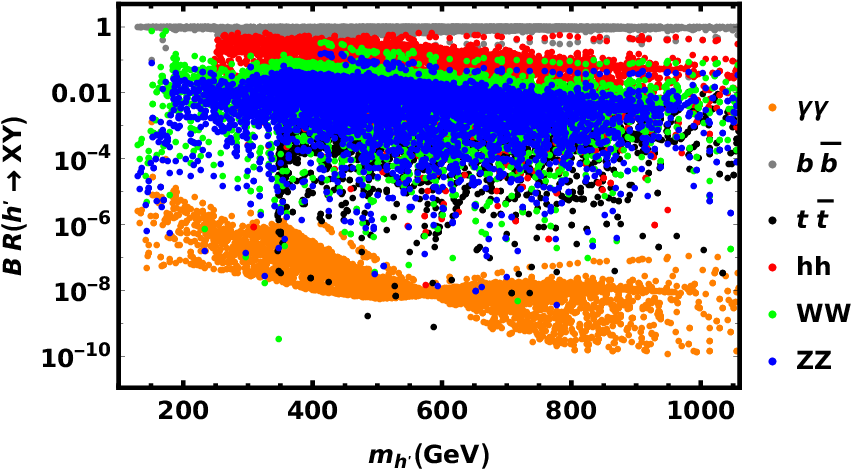}
\caption{\label{ggfusion}Left: The $h'$ production cross section from ggF as a function of its mass $m_{h'}$. The three benchmark points under consideration of Table~\ref{Bnchmrkpnts} are surrounded by blue circles.~Right: Branching ratios (BR) of $h'$ decays versus its mass $m_{h'}$. The relevant parameters values are used as in Fig.~\ref{massvspar}.}
\end{figure}

The heavy Higgs boson, $h'$, is mainly produced at the LHC from the gluon-gluon fusion (ggF) process, which induces about $90\%$ of its total production cross section at the LHC. Other production mechanisms, like vector boson fusion (VBF), Higgs strahlung and Higgs production from top fusion associated with top quark, represent the remaining ratio of the $h'$ production. In Fig.~\ref{ggfusion}~(left), we show the $h'$ ggF-production cross section versus $m_{h'}$ for the scanned values of parameters as in Eq.~(\ref{fixpar}) and its preceding paragraph. It is noticeable from Fig.~\ref{ggfusion}~(left) that the $h'$ ggF production cross section $\sigma(pp\to h')$ can be of an order $\gsim 2~{\text{pb}}$ for $m_{h'} \lsim 400~\text{GeV}$. In this regard, we consider $L_{\text{int}}=300~{\text{fb}}^{-1}$ for relatively light $h'$ and $L_{\text{int}}=3000~{\text{fb}}^{-1}$ for heavy $h'$ and in different decay channels, at $\sqrt{s}=14~\text{TeV}$.

We checked that all our benchmark points, given in Table~\ref{Bnchmrkpnts}, are validated to satisfy the usual \higgsbounds~and \higgssignals~limits confronted with the latest LHC data~\cite{Bechtle:2013wla,Bechtle:2013xfa}.
Recall that \higgsbounds~and \higgssignals~are testing Higgs sector both neutral and charged Higgs bosons of the model against the published exclusion bounds from Higgs searches at the LEP, Tevatron and LHC experiments. They are providing important tests for compatibility of any model beyond the SM. 

Indeed, many HEP computational tools are used through this work from building the model analytically the way to the numerical manipulation. The LRIS model was first implemented into the \sarah~package for building it, and it was then passed to \spheno~\cite{Staub:2013tta,Porod:2011nf} for numerical spectrum calculations. After that, the \ufo~model was used in \madgraph~\cite{Alwall:2011uj} for \montecarlo~events generation and matrix-element calculation. After that, \pythia~was also used to simulate the initial and final state radiation, fragmentation, and hadronisation effects~\cite{Sjostrand:2007gs}. For detector simulation, the \pythia~output was passed to \delphes~\cite{deFavereau:2013fsa}. Finally, for data analysis, we used \madanalysis~\cite{Conte:2012fm}. In Fig.~\ref{ggfusion}~(right), we show the relevant $h'$ decay branching ratios as functions of $m_{h'}$. It is remarkable to notice that for $m_{h'} \leq 600~\text{GeV}$, the $h'$ decay branching ratio to two SM Higgs boson is not small, mainly $\text{BR}(h'\to hh)\geq 10\%$, which gives a hope for probing this heavy Higgs boson through this channel.

\subsection{Search for $h'$ Higgs boson in $h'\to hh\to b\bar{b}\gamma\gamma$}

\begin{figure}[ht]
\centering
\includegraphics[width=10cm,height=6cm]{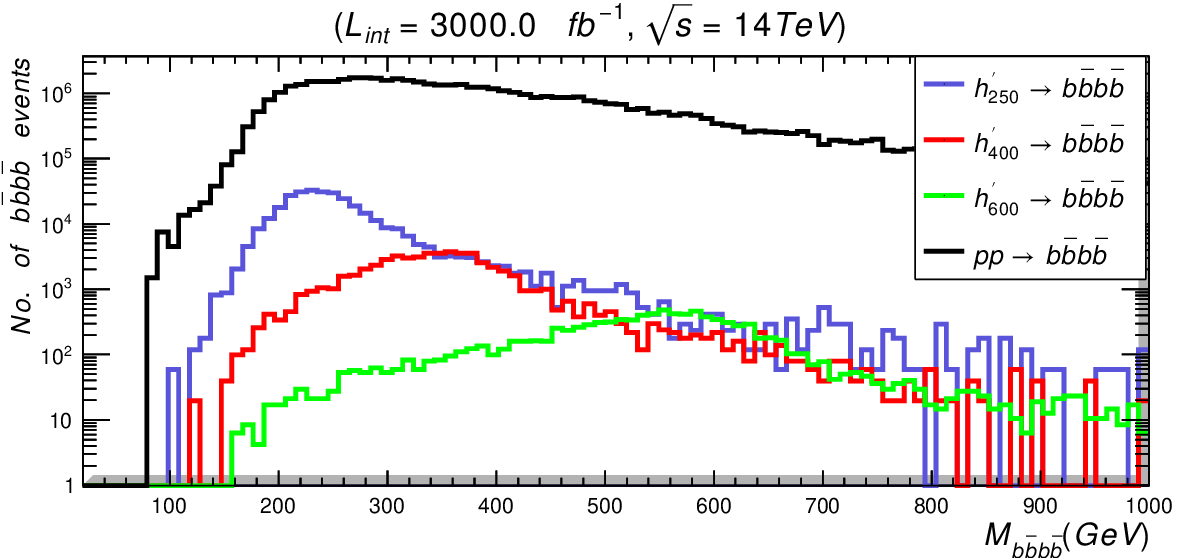}
\caption{\label{ppHhh4b}Number of signals events for $h'\to b\bar{b}b\bar{b}$ decays at mass $m_{h'}=250~\text{GeV (blue)}~,400~\text{GeV (red)}$, and $600~\text{GeV (green)}$ induced by the ggF versus the invariant mass of the final states $b\bar{b}b\bar{b}$, at $\sqrt{s}=14~\text{TeV}$ and $L_{\text{int}}=3000~{\text{fb}}^{-1}$ alongside the relevant background events (black) after applying the cut flow $|\eta|<2.4,~(P_T)_{bb}>30.0$ and $100.0<M_{bb}<150.0$.}
\end{figure}

We began with the decay $h'\to hh\to b\bar{b}b\bar{b}$ for probing $h'$ at the LHC, as the branching ratio $\text{BR}(h\to b\bar{b})$ is the largest of $h$ decays. However, this process has a huge background and the signal is much lower than the relevant background, even for a quite heavy Higgs boson: $m_{h'}>600~\text{GeV}$~\cite{Aaboud:2018knk,Sirunyan:2018tki} as in Fig.~\ref{ppHhh4b}. We found that no set of cuts can be used to increase the statistical significance of the signal enough to overcome the corresponding background, as in Fig.~\ref{ppHhh4b}. Specifically, we looked at the selection cuts for the following events on the pseudorapidity, the transverse momentum and the invariant mass of any combination of the four final states' $b$ jets, respectively: $|\eta_{bb}|<2.4,~(P_T)_{bb}>30.0~\text{GeV}$ and $100.0~\text{GeV}<M_{bb}<150.0~\text{GeV}$. Because our signal contains four final states $b$ jets from a pair of on shell SM Higgs bosons, it is expected that the majority of the signal events occur within a window centred on the SM Higgs mass $100.0~\text{GeV}<M_{bb}<150.0~\text{GeV}$. The corresponding relevant background events $b$-jets final states, on the other hand, are produced from a variety of other sources, the vast majority of which fall outside the above window. As a result, selecting events in this window is more likely to exclude events from the background than it is to exclude events from the signal. We examined all relevant signal and background kinematics and noticed that the used kinematics of $\eta_{bb},(P_T)_{bb},M_{bb}$ are the only discriminators that can be used to overweight signal over background and that our choices
optimized their usage based on cut efficiencies and the relative signal-to-background significance~$\text{S}/\sqrt{\text{B}}$. Also, we applied the cuts used in the experimental references~\cite{Aaboud:2018knk,Sirunyan:2018tki} on $|\eta_b|<2.5~(2.4)$ and $(P_T)_b>40~(30)~\text{GeV}$ of each single $b$ jet of the final states but we found no hope to enhance the significance and it gave results much less than we obtained by the cuts we used above. 
We also looked at the decay process $h'\to hh\to 2b+2W\to bb\ell\nu\ell\nu$, and we found that $\sigma(pp\to h'\to hh\to 2b+2W\to bb\ell\nu\ell\nu)\sim\mathcal{O}(10^{-15})~\text{pb}$. This unusually small cross section is caused by various
types of suppression for the corresponding amplitude. As a result, $h'$ can not be probed through this channel even at
high luminosity $L_{\text{int}}=300,3000~{\text{fb}}^{-1}$ and small $m_{h'}$~\cite{Sirunyan:2017guj,2020135145}. Therefore, in our analysis we are going to focus on the process $h'\to hh\to b\bar{b}\gamma\gamma$.

\begin{figure}[ht]
\centering
\includegraphics[scale=0.5]{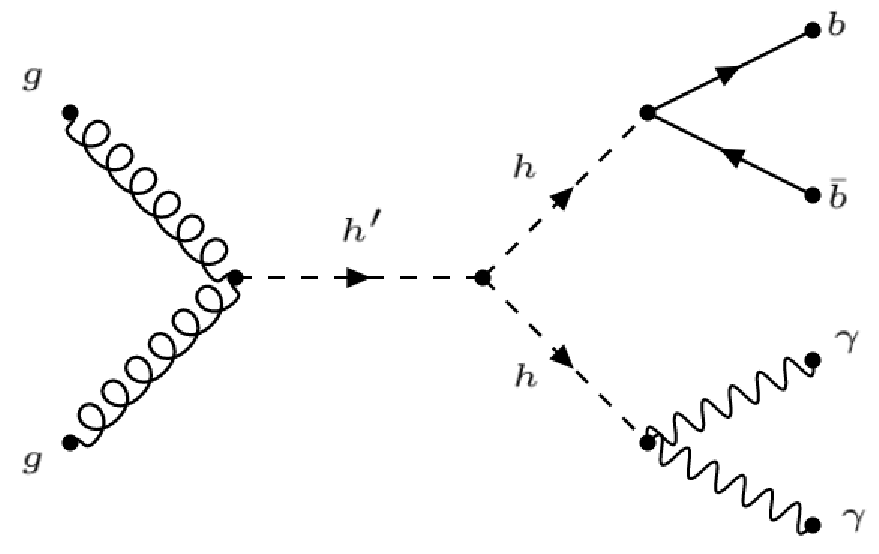}
\caption{\label{feynmanHhh}Feynman diagram for the $h'$ ggF production and decay process $ gg\to h'\to hh\to b\bar{b}\gamma\gamma$.}
\end{figure}

The on shell SM Higgs pair production from $h'$, followed by their decays $h'\to hh\to b\bar{b}\gamma\gamma$ is given by the Feynman diagram in Fig.~\ref{feynmanHhh}. As mentioned, here we adapt the following different values of $h'$-mass: $m_{h'}=250~\text{GeV},~400~\text{GeV}$, and $600~\text{GeV}$.

\begin{figure}[ht]
\centering
\includegraphics[scale=0.5,width=100mm]{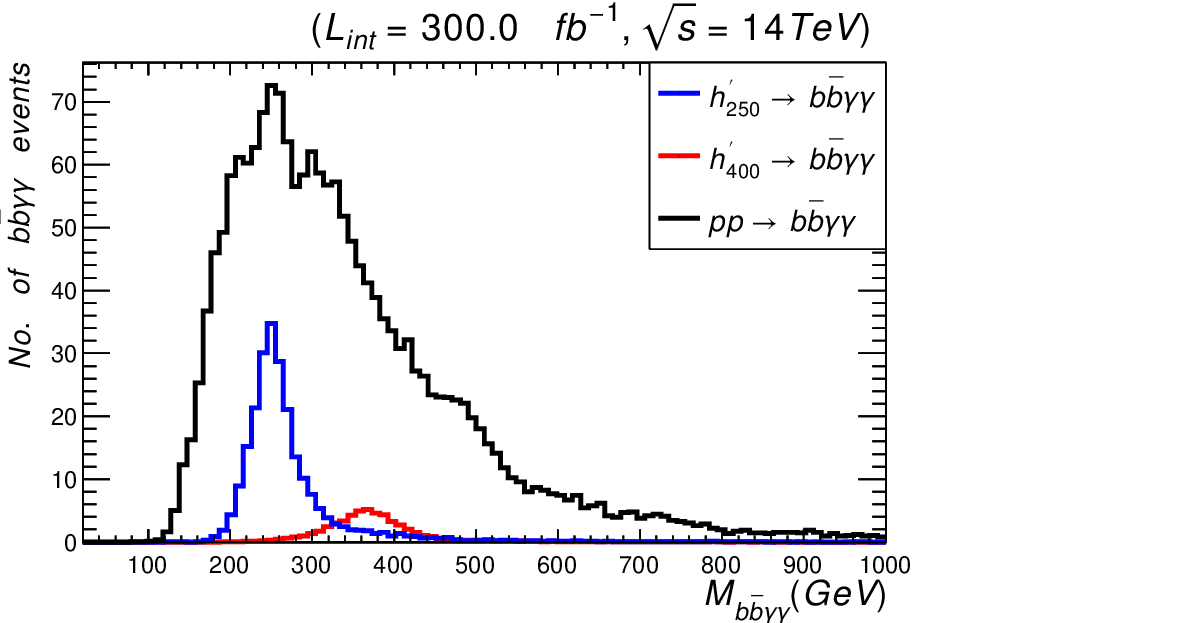}~\includegraphics[scale=0.5,width=100mm]{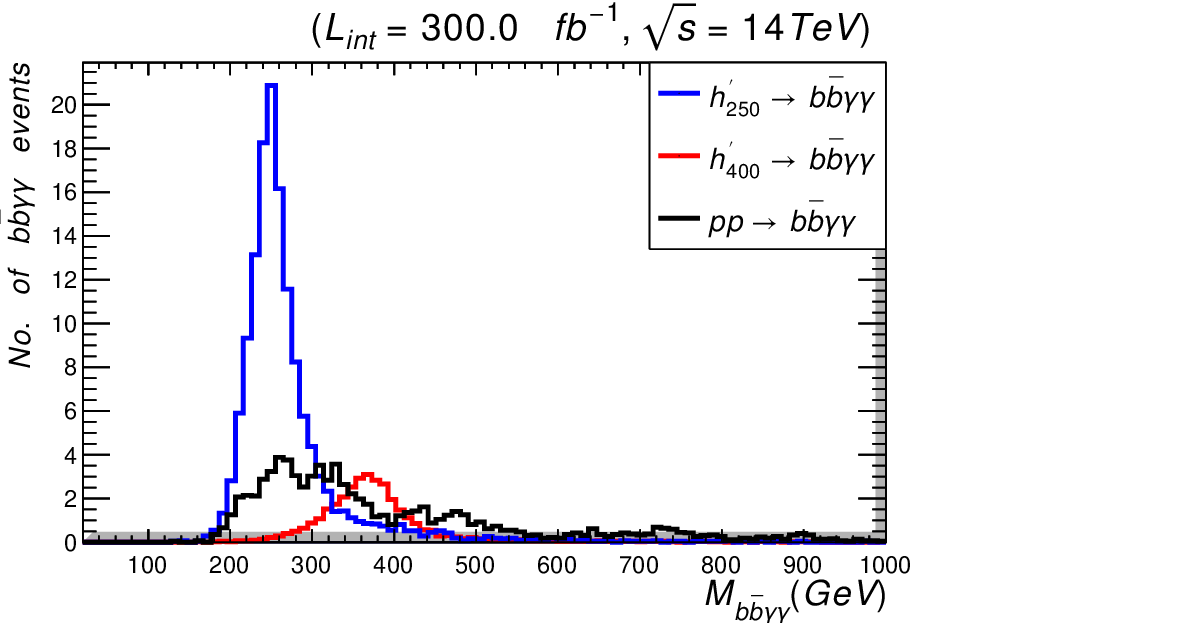}
\caption{\label{hp250400bkg}Number of signal events for $h'\to b\bar{b}\gamma\gamma$ decays at mass $m_{h'}=250~\text{GeV}$ (blue) and $400~\text{GeV}$ (red) induced by ggF versus the invariant mass of the final states $ b\bar{b}\gamma\gamma$, at $\sqrt{s}=14~\text{TeV}$ and $L_{\text{int}}=300~{\text{fb}}^{-1}$ alongside the relevant background events (black) before~(left) and after~(right) applying the cut flow of Table~\ref{cuttb300}. The corresponding values of cross sections and branching ratios are given in Table~\ref{xsecmhp}.}
\end{figure}

As the $h'$ decay width $\Gamma_{h'}$ is much smaller than its mass, $\Gamma_{h'}/m_{h'}\ll1$, the narrow width approximation can be used and the total cross section $\sigma(p p\to h'\to hh\to b\bar{b}\gamma\gamma)$ can be approximated as
\begin{equation}
\sigma(pp\to h'\to hh\to b\bar{b}\gamma\gamma ) \approx \sigma(pp\to h')\times \text{BR}(h'\to hh)\times \text{BR}(h\to b\bar{b})\times \text{BR}(h\to\gamma\gamma), 
\end{equation}
where the $h'\to hh$ decay branching ratio $\text{BR}(h'\to hh)$ is given in terms of the coupling $g_{h'hh}$ of Eq.~(\ref{hphh}). In Table~\ref{xsecmhp} the explicit values of the cross section and decay ratio of $h'$ are presented for the three values of $m_{h'}$, under consideration. 
\begin{center}
\begin{table}[ht]
\centering
\begin{tabular}{|c|c|c|c|}
\hline 
$m_{h'}~(\text{GeV})$ & $\sigma(pp\to h')~(\text{pb})$ & $\text{BR}(h'\to hh)$ & $\sigma(pp\to h'\to hh\to b\bar{b}\gamma\gamma )~(\text{fb})$ \\ \hline 
$250$ & $12.140$ & $0.30$ & $6.30$ \\ \hline 
$400$ & $5.050$ & $0.20$ & $1.01$ \\ \hline 
$600$ & $0.504$ & $0.18$ & $0.05$ \\ \hline 
\end{tabular}
\caption{\label{xsecmhp}$pp\to h'$ production cross section and its $h'\to hh$ decay branching ratio and the total cross section for its production and decay process $pp\to h'\to hh\to b\bar{b}\gamma\gamma$ for three different values of $m_{h'}=250~\text{GeV},~400~\text{GeV}$, and $600~\text{GeV}$.}
\end{table}
\end{center}
For potential discovery of $h'$ at the LHC, we analyze both its signal and the corresponding relevant background from the SM processes. There are many reducible background processes~\cite{Chang:2018uwu}
\begin{equation}
pp\to bbh\gamma\gamma/bbja/bbjj/cc\gamma\gamma/ccj\gamma/jj\gamma\gamma/ggh\gamma\gamma/tt/tt\gamma/tth\gamma\gamma/bbz\gamma\gamma/zh\gamma\gamma.
\end{equation}
The following preselection cuts at parton level are imposed in order to avoid any divergence in the parton-level calculations~\cite{ATLAS:2017muo,Chang:2018uwu}:
\begin{enumerate}
\item The pseudorapidity $\eta$ of the two photons must be in acceptance of the detector so $|\eta_{\gamma\gamma}|\leq 2.4$.
\item The pseudorapidity $\eta$ of the two jets must be in acceptance of the detector so $|\eta_{jj}|\leq 2.4$.
\item The transverse momentum $P_T$ of the two jets satisfies $(P_T)_{jj}\geq 20~\text{GeV}$.
\item The transverse momentum $P_T$ of the two photon satisfies $(P_T)_{\gamma\gamma}\geq 25~\text{GeV}$.
\end{enumerate}
All these backgrounds can be reduced by appropriate kinematical cuts on pseudorapidity $\eta_{ab}$, transverse momentum $(P_T)_{ab}$, invariant mass $M_{ab}$ of two final states objects $a,b$, and the angular distance $(\Delta R)_{ab}$ between $a,b$ in the transverse plane, as specified in the cut flow tables below. The most relevant background processes which compete with our signal are the irreducible ones, $pp\to b\bar{b}\gamma\gamma$ and $pp\to zh\to b\bar{b}\gamma\gamma$. The later one can be also reduced down by the same set of cuts which are used in Table~\ref{cuttb300}. 
\begin{table}[ht]
 \begin{center}
 \begin{tabular}{|c|c|c|c|}
 \hline
 Cuts (select) & Signal (S): $m_{h'}=250~\text{GeV}~(400~\text{GeV})$& Background (B) & S/$\sqrt{\text{B}}$ \\ \hline
 Initial (no cut) & $1904.00~(308.00)$ & $25058.00$ & $12.000~(1.950)$\\ \hline
$M_{\gamma\gamma} > 119.5~\text{GeV}$ & $846.70 \pm 21.70~(177.95 \pm 8.82)$ & $3015.10 \pm 51.50$ & $15.419 \pm 0.00527~(3.241 \pm 0.00272)$\\ \hline
 $M_{\gamma\gamma} < 130.5~\text{GeV}$ & $843.90 \pm 19.30~(175.80 \pm 8.36)$ & $766.40 \pm 19.20$ & $30.430 \pm 0.01500~(6.319 \pm 0.01540)$\\ \hline
 \end{tabular}
 \end{center}
 \caption{\label{cuttb300}Cut flow charts for the $h'\to hh\to b\bar{b}\gamma\gamma$ signal versus its relevant background and the corresponding number of events and significance at $300~{\text{fb}}^{-1}$ and $\sqrt{s}=14~\text{TeV}$ for $m_{h'}=250~\text{GeV}~(400~\text{GeV})$.}
\end{table}
In Fig.~\ref{hp250400bkg}, we show the number of signal events distributions for $m_{h'}=250~\text{GeV}$ and $400~\text{GeV}$ with the relevant irreducible background before~(left) and after~(right) applying cuts in Table~\ref{cuttb300}, respectively. 
In our signal, the final states kinematics are all boosted by the two on shell SM-like Higgs bosons and their distributions are narrowed and peaked around our $h'$ windows. This behavior of signal events is unlike that of the background events where final states have many sources and their kinematics' distributions are usually broadened throughout the whole range of analysis. For this, we notice the high relative reduction of the background to our signal when we restrict our analysis on events of kinematics which are expected to be distributed about our $h'$ as mentioned. We demand final state photons pairs of invariant masses as in Table~\ref{cuttb300} and this increases our signal significance in both cases of $m_{h'}=250~\text{GeV},~400~\text{GeV}$. This situation is shown in Fig.~\ref{hp250400bkg}, where the signals are first much less than the relevant background~(left) and choosing the certain cuts of Table~\ref{cuttb300} diminishes the background and makes the signals events finally dominate the background. The benchmark point with $m_{h'}=600~\text{GeV}$ is not included here as its cross section is quite tiny with the considered ($L_{\text{int}}=300~{\text{fb}}^{-1}$). 

The cut flow in Table~\ref{cuttb300} is chosen upon full analysis of final state signal and background kinematics to optimize relative signal-to-background significance ($\text{S}/\sqrt{\text{B}}$). As mentioned earlier, for $m_{h'}=250~\text{GeV},~400~\text{GeV}$ with $L_{\text{int}}=300~{\text{fb}}^{-1}$ the most relevant cuts were taken around the two SM Higgs peaks in the two photon and two jets invariant mass distributions. Applying these cuts, a major part of the background events were excluded, as it has broad distributions because they were generated as elastic scatering rather than being from resonances at the regions of interest. Eventually, the backgrounds reduced relative to the signals at $m_{h'}=250~\text{GeV},~400~\text{GeV}$ as in Fig.~\ref{hp250400bkg}~(right).

\begin{figure}[ht]
\centering
\includegraphics[scale=1.]{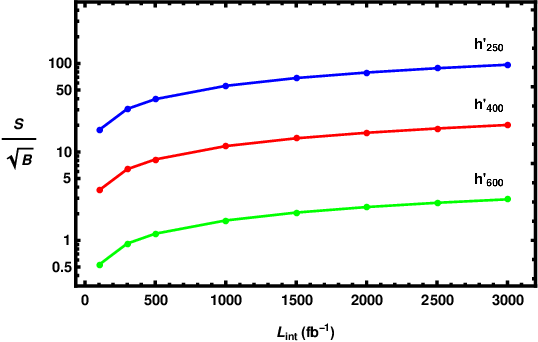}
\caption{\label{siglumihh}Significance of the $h'\to b\bar{b}\gamma\gamma$ signal of Fig.~\ref{hp250400bkg} relative to the corresponding background versus $L_{\text{int}}$ at mass $m_{h'}=250~\text{GeV}$ (blue), $400~\text{GeV}$ (red) and $600~\text{GeV}$ (green). Data are produced at $\sqrt{s}=14~\text{TeV}$, and points are interpolated between values of $L_{\text{int}}=100,300,500,1000,1500,2000,2500~{\text{fb}}^{-1}$ and $L_{\text{int}}=3000~{\text{fb}}^{-1}$. Notice that event rates are computed after the cuts described in Table~\ref{cuttb300}, and the relative significance of the signals increases with $L_{\text{int}}$.}
\end{figure}

From these results, it is clear that the SM-like Higgs boson pair production with $2\gamma+2b$-jets can be the smoking gun for probing the heavy $CP$-even neutral Higgs boson in this class of models that allows for a significant coupling between $h'$ and the SM-like Higgs $h$ boson, unlike several other extensions of the SM. The significance of the $h'\to b\bar{b}\gamma\gamma$ signal is presented in Fig.~\ref{siglumihh}, for $L_{\text{int}}$ values which vary from $100~{\text{fb}}^{-1}$ up to $3000~{\text{fb}}^{-1}$ at $\sqrt{s}=14~\text{TeV}$, for the usual three values of $m_{h'}$. It is clear that the signal significance increases with considered $L_{\text{int}}$ for each value of $m_{h'}$ giving better chances of $h'$ discovery with higher $L_{\text{int}}$.

\begin{table}[ht]
\begin{center}
\begin{tabular}{|c|c|c|c|}
\hline
Cuts (select)& Signal (S): $m_{h'}=600~\text{GeV}$ & Background (B) & S/$\sqrt{\text{B}}$\\\hline
Initial (no cut) & $155.000$ & $250650.00$ & $0.310$\\ \hline
$M_{bb} < 200.0~\text{GeV}$ & $52.250 \pm 5.18$ & $39823.60 \pm 82.40$ & $0.264 \pm 0.0008$ \\ \hline
$M_{\gamma\gamma} > 119.5~\text{GeV}$ & $34.436 \pm 5.91$ & $4252.00 \pm 64.70$ & $0.530 \pm 0.0010$\\ \hline
$M_{\gamma\gamma} < 130.5~\text{GeV}$ & $33.542 \pm 5.13$ & $1084.10 \pm 32.00$ & $1.018 \pm 0.0004$\\ \hline
$(\Delta R)_{\gamma\gamma} < 2.0$ & $29.230 \pm 4.35$ & $200.50 \pm 17.08$ & $1.062 \pm 0.0500$ \\\hline
$(\Delta R)_{bb} < 2.0$ & $27.680 \pm 4.46$ & $132.83 \pm 7.66$ & $2.409 \pm 0.0200$ \\ \hline
$(P_T)_{\gamma\gamma} > 200.0~\text{GeV}$ & $21.650 \pm 4.36$ & $57.65 \pm 7.70$ & $2.851 \pm 0.0260$ \\ \hline
\end{tabular}
\end{center} 
\caption{\label{cuttb600}Cut flow charts for the $h'\to hh\to b\bar{b}\gamma\gamma$ signal versus its relevant background and the corresponding number of events and significance at $L_{\text{int}}=3000~{\text{fb}}^{-1}$ and $\sqrt{s}=14~\text{TeV}$ for $m_{h'}=600~\text{GeV}$.}
\end{table}

\begin{figure}[H]
\centering
\includegraphics[scale=0.5]{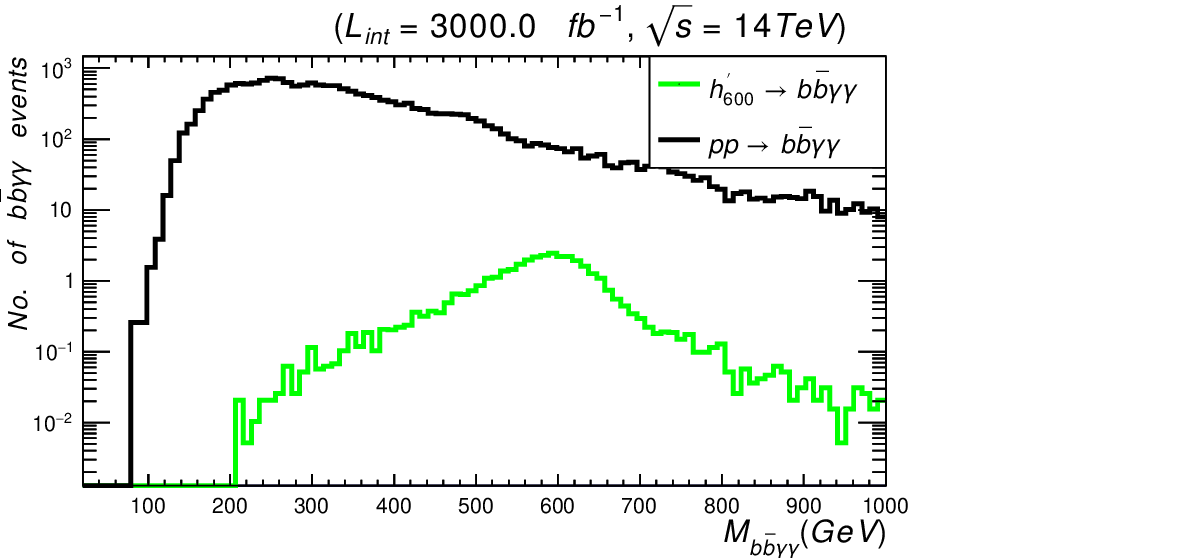}\includegraphics[scale=0.5]{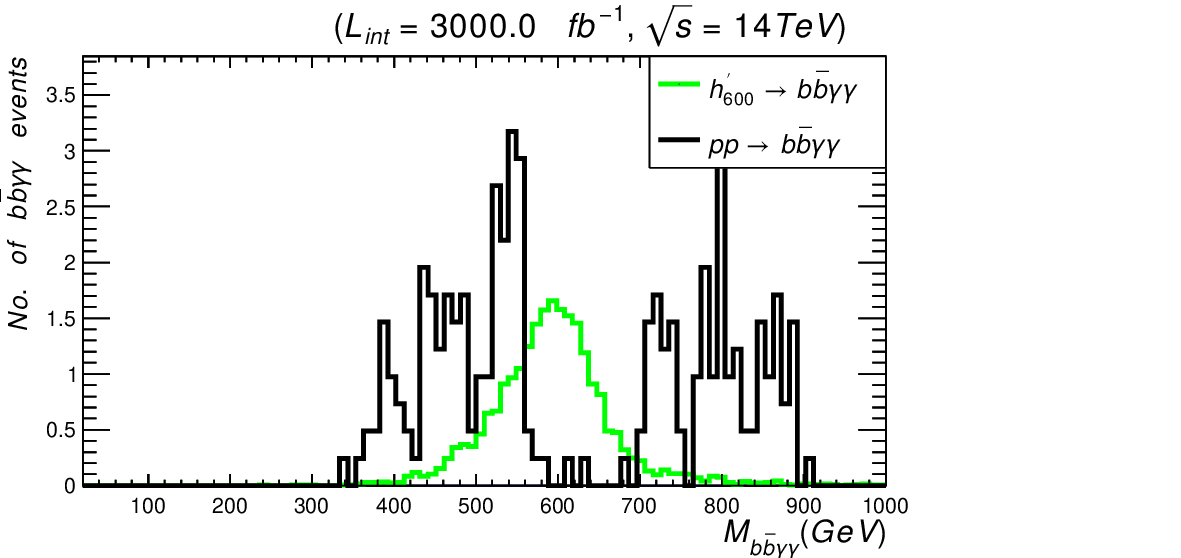}
\caption{\label{hp600bkg}Number of signal events for $h'\to b\bar{b}\gamma\gamma$ decays at mass $m_{h'}=600~\text{GeV}$ (green) induced by ggF versus the invariant mass of the final states $ b\bar{b}\gamma\gamma$, at $\sqrt{s}=14~\text{TeV}$ and $L_{\text{int}}=3000~{\text{fb}}^{-1}$ alongside the relevant background events background (black) before~(left) and after~(right) applying the cut flow set of Table~\ref{cuttb600}. The corresponding values of cross sections and branching ratios are given in Table~\ref{xsecmhp}. Left panel is plotted on log-scale vertical axis for the signal to show up relatively.}
\end{figure}

With $m_{h'}=600~\text{GeV}$, one must consider higher $L_{\text{int}}\sim 3000~{\text{fb}}^{-1}$, as the associated production and decay cross section are quite small. Here we apply the cut flow given in Table~\ref{cuttb600}. After applying all these cuts, the final distributions of this event are shown in Fig.~\ref{hp600bkg}. According to the plots shown in Fig.~\ref{siglumihh} and Fig.~\ref{hp600bkg}, it is clear that even $h'$ with $m_{h'}\gsim 600~\text{GeV}$ can still be discovered, but at the High-Luminosity Large Hadron Collider (HL-LHC), as it requires $L_{\text{int}}$ of order $L_{\text{int}}=3000~{\text{fb}}^{-1}$. 

\subsection{Search for $h'$ Higgs boson in $h'\to ZZ\to 4\ell$}
\begin{figure}[ht]
\centering
\includegraphics[scale=0.5]{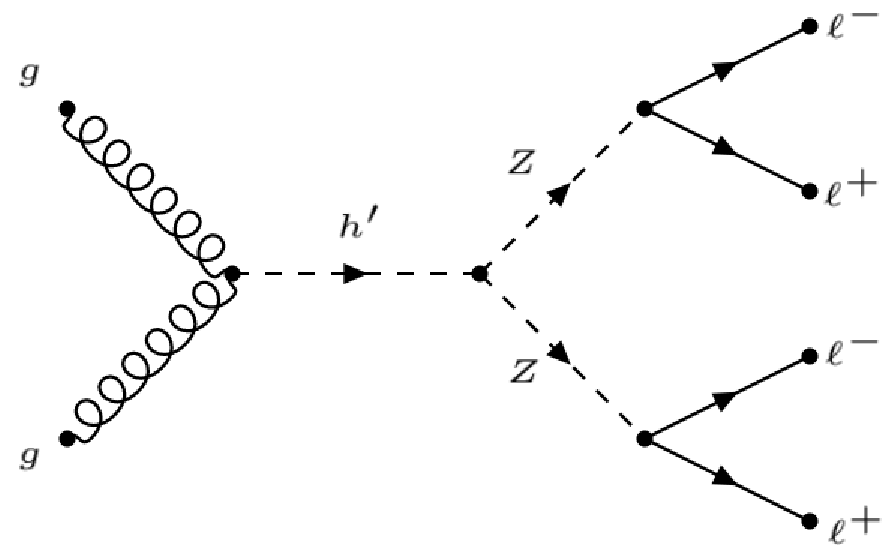}
\caption{\label{feynmanZZ}Feynman diagram for the $h'$ ggF production and decay process $ gg\to h'\to ZZ\to 4\ell$.}
\end{figure}

Here, we consider the possibility of probing $h'$ through its final decay into four charged leptons, along the process $pp\to h'\to ZZ\to 4\ell~(\ell=e,~\mu)$ with the Feynman diagram in Fig.~\ref{feynmanZZ}.

In the narrow width approximation, the total cross section can be written as
\begin{equation}
\sigma(pp\to h'\to ZZ\to 4\ell ) \approx \sigma(pp\to h')\times \text{BR}(h'\to ZZ)\times (\text{BR}(Z\to 2\ell))^2,
\end{equation}
and the $g_{h'ZZ}$ coupling is given in Eq.~(\ref{hpzz}). As explained below Eq.~(\ref{hpzz}), the $g_{h'ZZ}$ coupling can be as large as $\mathcal{O}(60)~\text{GeV}$ causing the total cross section $\sigma(pp\to h'\to ZZ\to 4\ell)$ to drop quickly with the propagator mass $m_{h'}$ without any compensation from elsewhere, unlike the previous case of the $h'\to hh$ decay where the drop of $\sigma(pp\to h'\to hh\to b\bar{b}\gamma\gamma)$ due to the propagator mass $m_{h'}$ can be partially compensated for from the coupling $g_{h'hh}$. Therefore, it is rather difficult to detect $h'$ through this channel for heavy $h'$. In this aspect, we will focus our analysis on $m_{h'}=250~\text{GeV}$.
The $h'$ production cross section, its decay branching ratio and the total cross section of $\sigma(pp\to h'\to ZZ\to 4\ell)$ are explicitly shown in Table~\ref{xsechpzzmhp}, for $m_{h'}=250~\text{GeV}$ and $400~\text{GeV}$.
\begin{center}
\begin{table}[ht]
\centering
\begin{tabular}{|c|c|c|c|}
\hline 
$m_{h'}~(\text{GeV})$ & $\sigma(pp\to h')~(\text{pb})$ & $\text{BR}(h'\to ZZ)$ & $\sigma(pp\to h'\to ZZ\to 4\ell)~(\text{fb})$ \\ 
\hline 
$250$ & $12.140$ & $0.050$ & $0.2428$ \\ 
\hline 
$400$ & $5.050$ & $0.025$ & $0.0579$ \\ 
\hline 
\end{tabular}
\caption{\label{xsechpzzmhp}$pp\to h'$ production cross section and its $h'\to ZZ$ decay branching ratio and the total cross section for its production and decay process $pp\to h'\to ZZ\to 4\ell$ for three different values of $m_{h'}=250~\text{GeV}$ and $400~\text{GeV}$.}
\end{table}
\end{center}

\begin{figure}[ht]
\centering
\includegraphics[scale=0.5]{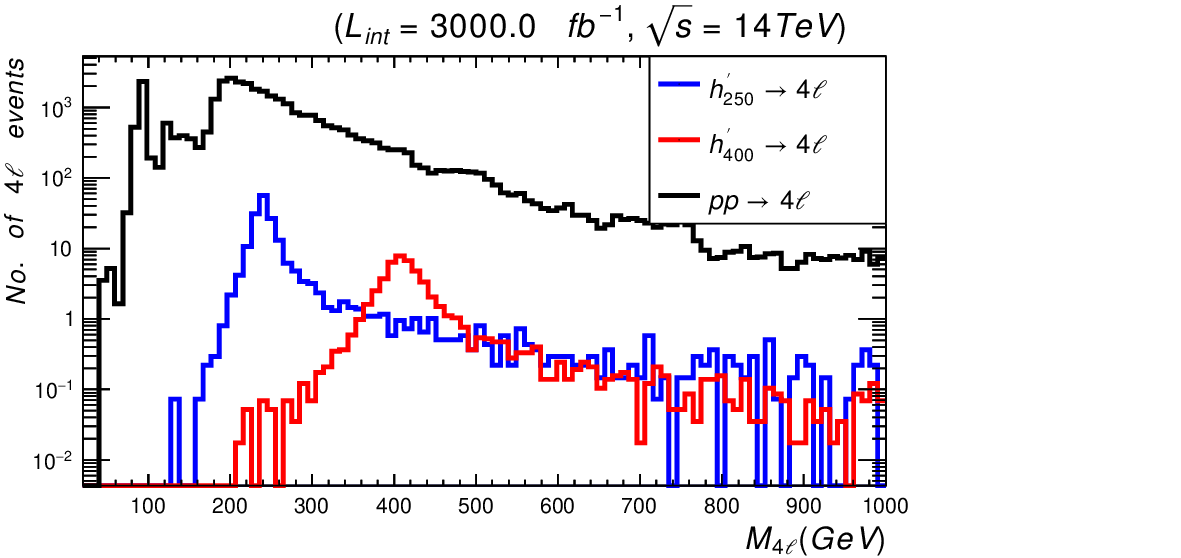}~\includegraphics[scale=0.5]{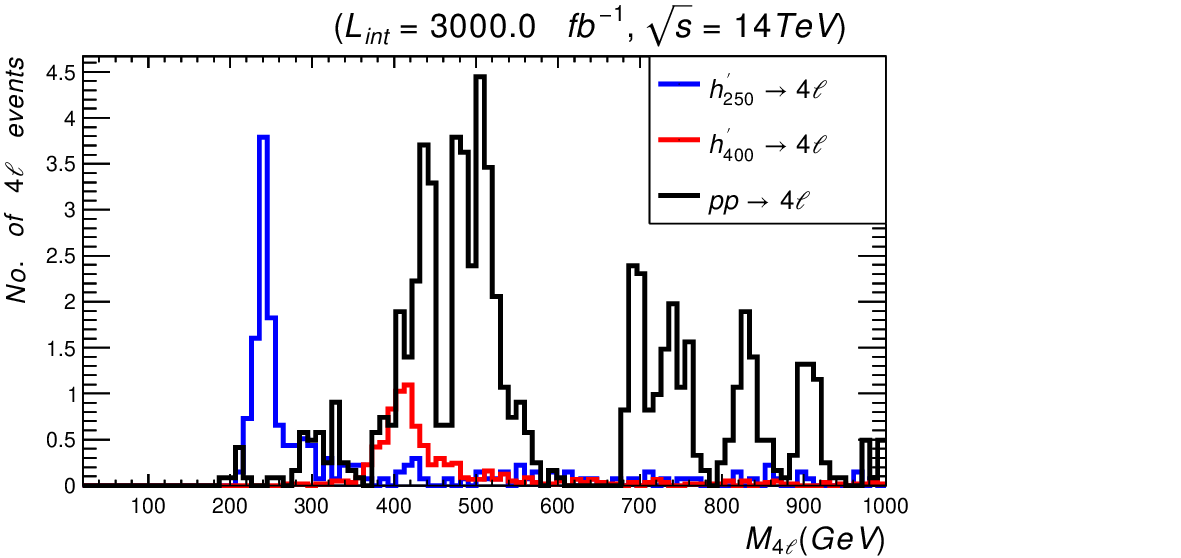}
\caption{\label{Hzzbcut}Number of signal events for $pp\to h'\to ZZ\to 4\ell$ decays at mass $m_{h'}=250~\text{GeV}$ (red) and $400~\text{GeV}$ (blue) induced by ggF versus the invariant mass of the final states $4\ell$, at $\sqrt{s}=14~\text{TeV}$ and $L_{\text{int}}=3000~{\text{fb}}^{-1}$ alongside the relevant background events background (black) before~(left) and after~(right) applying the cut flow of Table~\ref{cutHzz}. The corresponding values of cross sections and branching ratios are given in Table~\ref{xsechpzzmhp}.}
\end{figure}

It is clear from the results in Table~\ref{xsechpzzmhp} that with such small cross sections (fractions of fb), the number of associated events would be extremely smaller than the relevant background, as shown in Fig.~\ref{Hzzbcut}~(left). Here, we consider larger $L_{\text{int}}=3000~{\text{fb}}^{-1}$ for the both cases of $m_{h'}=250~\text{GeV}$ and $400~\text{GeV}$.

\begin{figure}[ht!]
\centering
\includegraphics[scale=1.]{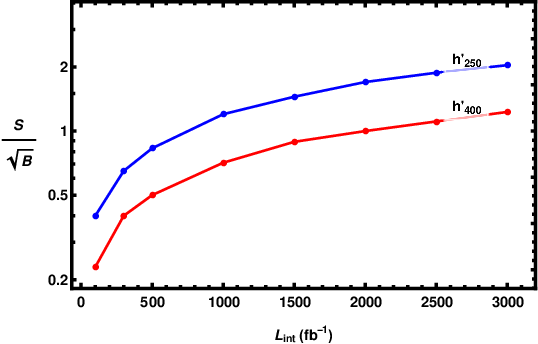}
\caption{\label{siglumizz}Significance of the $pp\to h'\to ZZ\to 4\ell$ signal of Fig.~\ref{Hzzbcut} relative to the corresponding background versus $L_{\text{int}}$ at mass $m_{h'}=250~\text{GeV}$ (blue) and $400~\text{GeV}$ (red). Data are produced at $\sqrt{s}=14~\text{TeV}$, and points correspond to $L_{\text{int}}=100,300,500,1000,1500,2000,2500~{\text{fb}}^{-1}$ and $L_{\text{int}}=3000~{\text{fb}}^{-1}$. Notice that event rates are computed after the cuts described in Table~\ref{cutHzz} and the relative significance of the signals increases with $L_{\text{int}}$.}
\end{figure}

Again, as our signal is boosted away by the high mass value of the $h'$ Higgs boson, an appropriate cut on the missing transverse hadronic energy $\slashed{H}_T=|-\sum_{\text{jet}}{(\vec{P}_T)_{\text{jet}}}|$ is applied as emphasized in Table~\ref{cutHzz} to enhance the relative significance of our signal to the corresponding irreducible background $pp\to 4\ell$.
\begin{table}[ht]
 \begin{center}
 \begin{tabular}{|c|c|c|c|}
 \hline
 Cuts (select) & Signal (S): $m_{h'}=250~\text{GeV}~(400~\text{GeV})$ & Background (B) & S/$\sqrt{\text{B}}$\\
 \hline
 Initial (no cut) & $728.00~(174.00)$& $79890.00$ & $2.58000~(0.43000)$ \\
 \hline
$\slashed{H}_T > 150.0~\text{GeV}$ & $58.65 \pm 7.34~(38.20 \pm 2.01)$ & $247.70 \pm 15.70$ & $2.02457 \pm 0.00790~(1.26340 \pm 0.00597)$ \\
\hline
 \end{tabular}
 \end{center}
 \caption{\label{cutHzz}Cut flow charts for the $h'\to ZZ\to 4\ell$ signal versus its relevant background and the corresponding number of events and significance at $3000~{\text{fb}}^{-1}$ and $\sqrt{s}=14~\text{TeV}$ for $m_{h'}=250~\text{GeV},~400~\text{GeV}$.}
\end{table}
Accordingly, the background is reduced significantly as shown in Fig.~\ref{Hzzbcut}~(right). However, the signal is also reduced and only a fraction of events would be available, which indicates that it is not possible to observe this signal via this channel.
In Fig.~\ref{siglumizz}, we present the significance of the $pp\to h'\to ZZ\to 4\ell$ signal to the corresponding background versus $L_{\text{int}}$, for mass $m_{h'}=250~\text{GeV}$ and $400~\text{GeV}$, and $\sqrt{s}=14~\text{TeV}$. It is clear from this plot that this signal can not be probed (\ie $\text{S}/\sqrt{\text{B}} > 1$), unless we have $L_{\text{int}}\sim 3000~{\text{fb}}^{-1}$. 

\section{\label{conc}Conclusions}
We have proposed a simplified LR model, where $SU(2)_R\times U(1)_{B-L}$ symmetry is broken spontaneously by the VEV of a scalar doublet $\chi_R$ around TeV scale, and the electroweak symmetry $SU(2)_L\times U(1)_Y$ is broken by the VEVs of two Higgs doublets merged from a single bidoublet $\phi$. We adopted IS mechanism to generate light neutrino masses. We also analyzed the Higgs sector in detail, in particular the three neutral $CP$-even Higgs bosons. We showed that the lightest of these particle can be assigned to the SM-like Higgs boson, with mass equals to $125~\text{GeV}$. The next lightest Higgs boson, $h'$, which is stemmed from the bidoublet neutral component is of order a few hundred~\text{GeVs}. 
We studied the LHC potential discovery for $h'$ in this class of models. We performed analysis for searches for $h'$ by looking for resonant peaks in the following two processes: $h'\to hh\to b\bar{b}\gamma\gamma$ and $h'\to ZZ\to 4\ell~(\ell=e,\mu)$. We considered three benchmark points, with $m_{h'}=250~\text{GeV},~400~\text{GeV}$, and $600~\text{GeV}$, at $\sqrt{s}=14~\text{TeV}$ and $L_{\text{int}}=300~{\text{fb}}^{-1}$ and $L_{\text{int}}=3000~{\text{fb}}^{-1}$. We emphasized that $h'$ can be probed with good statistical significances in di-Higgs channel, with $2\gamma+2b$-jets final states. While the channel of $Z$-pair production and decays to $4\ell$ is much less significant, it may be observed only at very high $L_{\text{int}}=3000~{\text{fb}}^{-1}$ and for light $h'$ with mass less than $300~\text{GeV}$.

\section*{acknowledgements}
{\it M. Ashry} and {\it K. Ezzat} would like to thank {\it A. Hammad} for the fruitful discussions. This paper is based upon work supported by Science, Technology $\&$ Innovation Funding Authority (STDF) under grant number 37272.

\section{Appendix}
\subsection{\label{apptdplmin}TADPOLE EQUATIONS AND POTENTIAL MINIMIZATION}
The minimum of the scalar potential (\ref{scalarpot}) is
\begin{align}
\langle V\rangle=V(\langle\phi\rangle,\langle\chi_R\rangle)
&=\frac{1}{4} \Big[\lambda_1(k_1^4+k_2^4)+4 \lambda_4 k_1 k_2(k_1^2+k_2^2)+2(\lambda_1+2 \lambda_{23})k_1^2 k_2^2+2\mu_1(k_1^2+k_2^2)+(\alpha_{13} k_1^2+\alpha_{12} k_2^2) v_R^2\nonumber\\
&+2 k_1 k_2 (4 \mu_2+\alpha_4 v_R^2)+2 \mu_3 v_R^2+\rho_1 v_R^4\Big],
\end{align}
where the VEVs satisfy the following tadpole equations 
\begin{align}
\frac{\partial\langle V\rangle}{\partial k_1}&=\lambda_1 k_1^3+\lambda_4 k_2(3 k_1^2+k_2^2)+k_1 \{k_2^2 (\lambda_1+2 \lambda_{23})+\mu_1+\frac{1}{2}\alpha_{13} v_R^2\}+2 k_2 \mu_2+\frac{1}{2} \alpha_4 k_2 v_R^2=0,\\
\frac{\partial\langle V\rangle}{\partial k_2}&=\lambda_1 k_2^3+\lambda_4 k_1(k_1^2+3 k_2^2)+k_2 \{k_1^2 (\lambda_1+2 \lambda_{23})+\mu_1+\frac{1}{2}\alpha_{12} v_R^2\}+2 k_1 \mu_2+\frac{1}{2} \alpha_4 k_1 v_R^2=0,\\
\frac{\partial\langle V\rangle}{\partial v_R}&=\frac{1}{2} v_R \{\alpha_{13} k_1^2+2 \alpha_4 k_1 k_2+\alpha_{12} k_2^2+2 (\mu_3+\rho_1 v_R^2)\}=0.
\end{align}
We solve them for $\mu_1,~\mu_2$ and $\mu_3$ as follows:
\begin{align}
\mu_1&=-\lambda_1 (k_1^2+k_2^2)-2\lambda_4 k_1 k_2-\frac{\alpha_{12}k_2^2-\alpha_{13}k_1^2}{2(k_2^2-k_1^2)}v_R^2,\\
\mu_2&=-\frac{1}{2}\lambda_4(k_1^2+k_2^2)-\lambda_{23} k_1 k_2-\frac{1}{4}(\alpha_4+\frac{\alpha_{32}k_1k_2}{k_2^2-k_1^2})v_R^2,\\
\mu_3&=-\frac{1}{2}(\alpha_{13} k_1^2+2 \alpha_4 k_1 k_2+\alpha_{12} k_2^2+2\rho_1 v_R^2).
\end{align}
where we define $\alpha_{1i}=\alpha_1+\alpha_i,~i=2,3$, $\alpha_{32}=\alpha_3-\alpha_2$ and $\lambda_{23}=2\lambda_2+\lambda_3$. 
\subsection{\label{appcopbd}COPOSITIVITY CONDITIONS OF THE HIGGS POTENTIAL}
To study the boundedness from below and hence, the stability, of the scalar potential (\ref{scalarpot}) we use the copositvity theorems of~\cite{Ping1993109,Kannike:2012pe} and follow the procedure used in~\cite{Ashry:2013loa} to ensure that the following symmetric matrix of the quartic terms, which are dominant at higher values of the fields, is copositive:
\begin{equation}
\begin{pmatrix}
\lambda_1 & \lambda_1 & \lambda_1+2\lambda_{23} & \lambda_1 & \frac{1}{2}\alpha_{13} & \frac{1}{2}\alpha_{12} \\
. & \lambda_1 & \lambda_1 & \lambda_1+2\lambda_3 & \frac{1}{2}\alpha_{12} & \frac{1}{2}\alpha_{13} \\
. & . & \lambda_1 & \lambda_1 & \frac{1}{2}\alpha_{12} & \frac{1}{2}\alpha_{13} \\
. & . & . & \lambda_1 & \frac{1}{2}\alpha_{13} & \frac{1}{2}\alpha_{12} \\
. & . & . & . & \rho_1 & \rho_1 \\
. & . & . & . & . & \rho_1
\end{pmatrix}.
\end{equation}
Copositivity of this matrix demands that $\lambda_1>0,~\rho_1>0$, and either of the following cases
\begin{enumerate}
\item $\lambda_1+2\lambda_{23}>0,~\lambda_1+2\lambda_3 >0,~\alpha_{12}>0,~\alpha_{13}>0$.
\item If $\lambda_1+2\lambda_{23}>0,~\lambda_1+2\lambda_3 <0,~\alpha_{12}>0,~\alpha_{13}>0$, then $\lambda_3 <0$ or $\lambda_1+\lambda_3 > 0$ and $\lambda_2 > 0$.
\item If $\lambda_1+2\lambda_{23}<0,~\lambda_1+2\lambda_3 >0,~\alpha_{12}>0,~\alpha_{13}>0$ then $\lambda_3 <0$,~$\lambda_2 < 0$ and $\lambda_1+\lambda_{23} >0$.
\item If $\lambda_1+2\lambda_{23}<0,~\lambda_1+2\lambda_3 <0,~\alpha_{12}>0,~\alpha_{13}>0$, then $\lambda_3 <0$.
\item If $\lambda_1+2\lambda_{23}>0,~\lambda_1+2\lambda_3 >0,~\alpha_{12}<0,~\alpha_{13}>0$, then $\lambda_1 \rho_1 > \frac{1}{4}\alpha_{13}^2$ and $\lambda_1 \rho_1 > \frac{1}{2}(\alpha_{12}+\alpha_{13})^2$.
\item If $\lambda_1+2\lambda_{23}>0,~\lambda_1+2\lambda_3 >0,~\alpha_{12}>0,~\alpha_{13}<0$, then $\lambda_1 \rho_1 > \frac{1}{4}\alpha_{12}^2$ and $\lambda_1 \rho_1 > \frac{1}{2}(\alpha_{12}+\alpha_{13})^2$.
\item If $\lambda_1+2\lambda_{23}>0,~\lambda_1+2\lambda_3 >0,~\alpha_{12}<0,~\alpha_{13}<0$,
\\then $\lambda_1 \rho_1 > \frac{1}{4}\alpha_{12}^2$ and $\lambda_1 \rho_1 > \frac{1}{4}\alpha_{13}^2$ or $\lambda_1 \rho_1 < \frac{1}{4}\alpha_{12}^2$ and $\lambda_1 \rho_1 < \frac{1}{4}\alpha_{13}^2$ .
\item If $\lambda_1+2\lambda_{23}<0,~\lambda_1+2\lambda_3 <0,~\alpha_{12}<0,~\alpha_{13}<0$,
\\then $\lambda_3< 0$ $\lambda_1 \rho_1 > \frac{1}{4}\alpha_{12}^2$ and $\lambda_1 \rho_1 > \frac{1}{4}\alpha_{13}^2$ or $\lambda_1 \rho_1 < \frac{1}{4}\alpha_{12}^2$ and $\lambda_1 \rho_1 < \frac{1}{4}\alpha_{13}^2$ .
\end{enumerate}
Finally, field redefinition could be done to make quartic terms in potential like $\phi_1^{0}\phi_2^{0}\phi_R^{+}\phi_R^{-}$ nonnegative definite again as in~\cite{Ashry:2013loa}.
\subsection{\label{appdet}Neutral $CP$-even Higgs Rotations and Determinant}
\noindent
The explicit rotation coefficients of the $CP$-even Higgs mass matrix (\ref{cpevenhgsms}) are~\cite{86718}
\begin{align}
Z^H_{11}&=\frac{f_{11}}{\sqrt{f_{11}^2+f_{21}^2+1}},\\
Z^H_{12}&=\frac{f_{21}}{\sqrt{f_{11}^2+f_{21}^2+1}},\\
Z^H_{13}&=\frac{1}{\sqrt{f_{11}^2+f_{21}^2+1}},\\
Z^H_{21}&=\frac{f_{12}(1+f_{21}^2)-f_{11}(1+f_{21}f_{22})}{\sqrt{(1+f_{11}^2+f_{21}^2)\{(f_{11}-f_{12})^2+(f_{21}-f_{22})^2+(f_{12}f_{21}-f_{11}f_{22})^2\}}},\\
Z^H_{22}&=\frac{f_{22}(1+f_{11}^2)-f_{21}(1+f_{11} f_{12})}{\sqrt{(1+f_{11}^2+f_{21}^2)\{(f_{11}-f_{12})^2+(f_{21}-f_{22})^2+(f_{12}f_{21}-f_{11}f_{22})^2\}}},\\
Z^H_{23}&=\frac{f_{11}(f_{11}-f_{12})+f_{21}(f_{21}-f_{22})}{\sqrt{(1+f_{11}^2+f_{21}^2)\{(f_{11}-f_{12})^2+(f_{21}-f_{22})^2+(f_{12}f_{21}-f_{11}f_{22})^2\}}},\\
Z^H_{31}&=\frac{(\text{sgn})(f_{22}-f_{21})}{\sqrt{(f_{11}-f_{12})^2+(f_{21}-f_{22})^2+(f_{12}f_{21}-f_{11}f_{22})^2}},\\
Z^H_{32}&=\frac{(\text{sgn})(f_{11}-f_{12})}{\sqrt{(f_{11}-f_{12})^2+(f_{21}-f_{22})^2+(f_{12}f_{21}-f_{11}f_{22})^2}},\\
Z^H_{33}&=\frac{(\text{sgn})(f_{12}f_{21}-f_{11} f_{22})}{\sqrt{(f_{11}-f_{12})^2+(f_{21}-f_{22})^2+(f_{12}f_{21}-f_{11}f_{22})^2}},
\end{align}
where the sign term is
\be
\text{sgn}=\text{sign}\{f_{11}(f_{23}-f_{22})+f_{12}(f_{21}-f_{23})+f_{13}(f_{22}-f_{21})\},
\ee
and $f_{ij}=f_i(m_{H_j}^2),~(i=1,2,~j=1\ldots3)$ and the functions $f_i$'s are
\begin{align}
f_1(x)&=~\frac{(m_{22}-x)(m_{33}-x)-m_{23}^2}{m_{12}m_{23}-m_{13}(m_{22}-x)},\\
f_2(x)&=-\frac{m_{12}(m_{33}-x)-m_{13}m_{23}}{m_{12}m_{23}-m_{13}(m_{22}-x)}.
\end{align}
The determinant of the $CP$-even Higgs mass matrix (\ref{cpevenhgsms}) is given by
\begin{align}
D^h&=v^2 v_R^2 (-(\alpha_{12} c_{\beta}+\alpha_{4} s_{\beta}) ((\alpha_{12} c_{\beta}+\alpha_{4} s_{\beta}) (v^2 (c_{2\beta} (\lambda_{23}-\lambda_1)+2 \lambda_{4} s_{2\beta})-\frac{1}{4} \alpha_{32} v_R^2 sc_{2\beta}+v^2 (\lambda_1+\lambda_{23})-\frac{\alpha_{32} v_R^2}{4})\nonumber\\
&-\frac{1}{4} (\alpha_{13} s_{\beta}+\alpha_{4} c_{\beta}) (4 v^2 (s_{2\beta} (\lambda_1+\lambda_{23})+2 \lambda_{4})+\alpha_{32} v_R^2 t_{2\beta}))+(\alpha_{13} s_{\beta}+\alpha_{4} c_{\beta}) ((\alpha_{12} c_{\beta}+\alpha_{4} s_{\beta}) (v^2 s_{2\beta} (\lambda_1+\lambda_{23})\nonumber\\
&+\frac{1}{4} \alpha_{32} v_R^2 t_{2\beta}+2 \lambda_{4} v^2)-(\alpha_{13} s_{\beta}+\alpha_{4} c_{\beta}) (v^2 (c_{2\beta} (\lambda_1-\lambda_{23})+2 \lambda_{4} s_{2\beta})-\frac{1}{4} \alpha_{32} v_R^2 sc_{2\beta}+v^2 (\lambda_1+\lambda_{23})+\frac{\alpha_{32} v_R^2}{4}))\nonumber\\
&+\rho_1 sc_{2\beta} (6 v^2 c_{2\beta} (\lambda_1 \lambda_{23}-\lambda_{4}^2)+2 v^2 c_{6\beta} (\lambda_1 \lambda_{23}-\lambda_{4}^2)+\alpha_{32} \lambda_{23} v_R^2 c_{4\beta}-4 \alpha_{32} \lambda_{4} v_R^2 s_{2\beta}-\alpha_{32} v_R^2 (2 \lambda_1+\lambda_{23}))).
\end{align}
\bibliographystyle{apsrev}\bibliography{Bib}
\end{document}